\documentclass{aa}
\usepackage{graphicx}
\usepackage[round,authoryear]{natbib}
\usepackage{amssymb}
\usepackage{amsmath}
\usepackage{multirow}
\usepackage{rotating}
\usepackage{bm}

\def\simlt{\lower.5ex\hbox{$\; \buildrel < \over \sim \;$}}
\def\simgt{\lower.5ex\hbox{$\; \buildrel > \over \sim \;$}}
\def\ltsim{\raise 2pt \hbox {$<$} \kern-1.1em \lower 4pt \hbox {$\sim$}}
\def\ltapprox{\raise 2pt \hbox {$<$} \kern-1.1em \lower 5pt \hbox {$\approx
$}}
\def\gtsim{\raise 2pt \hbox {$>$} \kern-1.1em \lower 4pt \hbox {$\sim$}}
\def\gtapprox{\raise 2pt \hbox {$>$} \kern-1.1em \lower 5pt \hbox {$\approx
$}}

\begin{document}


\title{Cosmic dance in the Shapley Concentration Core -- I. A study of the radio emission of the BCGs and tailed radio galaxies}

\author{G. Di Gennaro\inst{1,2,3}, 
		T. Venturi\inst{1}, 
		D. Dallacasa\inst{1,4}, 
		S. Giacintucci\inst{6}, 
		P. Merluzzi\inst{7},  
		G. Busarello\inst{7}, 
		A. Mercurio\inst{7}, S. 
		Bardelli\inst{5}, 
		F. Gastaldello\inst{8}, 
		A. Grado\inst{7}, 
		C. P. Haines\inst{9}, 
		L. Limatola\inst{7}, 
		M. Rossetti\inst{8}}

\institute
{
INAF -- Istituto di Radioastronomia, via Gobetti 101, I-40129, Bologna, Italy
\and
Harvard-Smithsonian Center for Astrophysics, 60 Garden Street, Cambridge, MA 02138, USA
\and
Leiden Observatory, Leiden University, PO Box 9513, 2300 RA Leiden, The Netherlands
\and
Dipartimento di Fisica e Astronomia, Universit{\`a} di Bologna, via Gobetti 93/2, I-40129 Bologna, Italy
\and
INAF-Astrophysics and Space Science Observatory Bologna, via Gobetti 93/3, I-40129 Bologna, Italy
\and
Naval Research Laboratory, 4555 Overlook AvenueSW, Code 7213, Washington, DC 20375, USA
\and
INAF -- Osservatorio Astronomico di Capodimonte, Salita Moiariello 16, 80131 Napoli, Italy
\and
INAF -- IASF Milano, via Bassini 15, I-20133 Milano, Italy
\and
INAF -- Osservatorio Astronomico di Brera, via Brera 28, 20121, Milano, Italy
}



\abstract
{The Shapley Concentration ($\langle z \rangle\approx0.048$) covers several degrees
in the Southern Hemisphere, and includes galaxy clusters in 
advanced evolutionary stage,  groups of clusters in the very early stages of merger, fairly massive clusters with ongoing 
accretion activity, and smaller groups located in filaments in the regions between the main clusters.} 
{With the goal to investigate the role of cluster mergers and accretion on the radio galaxy population, we performed 
a multi-wavelength study of the brightest cluster galaxies (BCGs) and of the galaxies showing extended radio emission 
in the cluster complexes of Abell 3528 and Abell 3558. In total, our study is based on a sample of 12 galaxies.}
{We observed the clusters with the Giant Metrewave Radio Telescope (GMRT) at 235, 325 and 610 MHz, 
and with the Very Large Array (VLA) at 8.46 GHz. We complemented
our study with the TIFR GMRT Sky Survey (TGSS) at 150 MHz, the Sydney University Molonglo Sky Survey (SUMSS) at 843 MHz
 and the Australia Telescope Compact array (ATCA) at  1380, 1400, 2380, 
and 4790 MHz data. Finally, optical imaging with ESO-VST is also available for the host galaxies as well as the mid-infrared coverage with WISE.}
{We found deep differences in the properties of the radio emission of the BCGs in the two cluster complexes. The BCGs in the A\,3528 complex and in A\,3556, which are relaxed cool-core objects, are powerful active radio galaxies. They also present hints of restarted activity. On the contrary, the BCGs in A\,3558 and A\,3562, which are well known merging systems, are very faint, or quiet, in the radio band.
The optical and infrared properties of the galaxies, on the other hand, are fairly similar in the two complexes, showing all passive red galaxies. }
{Our study shows remarkable differences in the radio properties of the BGCs, which we relate to the different dynamical state of the host cluster. On the contrary, the lack of changes between such different environments in the optical band suggest that the dynamical state of galaxy clusters does not affect the optical counterparts of the radio galaxies, at least over the life-time of the radio emission.}
{}

\keywords{radio continuum: galaxies - galaxies: clusters: general - galaxies: clusters: individual: A3528  - galaxies: clusters: individual: A3532 - galaxies: clusters: individual: A3556 - galaxies: clusters: individual: A3558 - galaxies: clusters: individual:
A3562}

\titlerunning{Cosmic dance in the Shapley Concentration Core}
\authorrunning{G. Di Gennaro et al.}
\maketitle

\section{Introduction}\label{sec:intro}
Cluster mergers are the most energetic phenomena in the Universe, releasing energies of the order of 
$10^{\rm 63-64}$ erg depending on the mass of the clusters and on the relative velocity of the merging halos. 
In a hierarchical cold dark 
matter scenario,  these phenomena are the natural way to form rich clusters of galaxies, but it is still unknown how 
such amounts of gravitational energy are dissipated and which are the consequences on the environment 
(intracluster medium, magnetic fields and in-situ relativistic particles) and on the galaxy population.

Mergers and group accretion leave important footprints in the radio band. 
Beyond the well known diffuse cluster sources in the form of halos, relics and mini-halos, whose relation to
the cluster dynamical state is an established result from both the observational and theoretical point of view \citep[e.g.,][]{markevitch+vikhlinin07,venturi+08,vanweeren+11,zuhone+13,cassano+13,brunetti+14,giacintucci+14b,giacintucci+17}, further
crucial information on the cluster dynamics and on
the interaction between the radio plasma and the ICM comes from the galaxy population. 
Double radio galaxies in galaxy clusters (mainly FRIs and FRI/II transition objects, referenze) often show
 bent and distorted jets and lobes, which are usually classified on the basis of the angle 
between the two jets \citep{miley80}. 
Head-tailed (HT) and narrow-angle tailed (NAT) radio galaxies (sometimes called also C-shaped sources) are associated 
with  non-dominant cluster galaxies, and are interpreted as the result of ram pressure exerted by the intergalactic 
medium on the double-sided radio emission \citep{odea+85}. 
Wide-angle tailed (WAT) radio galaxies, on the other hand are often associated with the central brightest cluster
galaxies (BCGs) and their 
shape is explained as a combination of bulk motion of the ICM as consequence of clusters mergers and ``cluster
weather'' (\citealt{burns+98} and see \citealt{feretti+venturi02} for a review).

BCGs are the most massive and luminous elliptical galaxies in the Universe. They are located in the proximity of the bottom of the cluster potential well, and are usually active in the radio band: the fraction of radio loud BCGs is much higher than for the other elliptical galaxies, reaching up to 30\% for 
$L_{\rm 1.4~GHz}>10^{23}$ W Hz$^{-1}$, and the normalization of their radio luminosity function is the highest among elliptical galaxies \citep{best+07}.  
Moreover, the radio properties of these radio galaxies are strongly dependent on the central properties of the host cluster,
both in the local Universe and at intermediate redshift. The probability to find a radio loud BCG is much higher in relaxed clusters, and such probability increases if we are looking at ``strong'' cool--core clusters \citep{cavagnolo+08,sun09,kale+15}.
The AGN feedback of the BCGs is also an established result: the nuclear radio emission may take the form of mechanical
feedback to the ICM, preventing gas cooling in dynamically relaxed environments \citep[see][]{mcnamara+07,mcnamara+12a}. 

The galaxy environment is a determinant for both the
morphology-density \citep{dressler80,dressler+97} and the star
formation-density \citep{butcher+oemler84,lewis+02,kauffman+04}
 relations observed at redshift $z\sim0$. These relations
show that late-type, blue, star-forming galaxies are predominant in
the field, while early-type, red, passive galaxies are preferentially
found in galaxy clusters. This suggests that blue galaxies accreted
from the field in the past have been transformed into the passive lenticular
and dwarf ellipticals found in local clusters. 
The proposed and investigated mechanisms affecting the galaxy
properties include gravitational and tidal interactions amongst
galaxies \citep{toomre+toomre72,moore+96}, between galaxies
and the cluster gravitational field \citep{byrd+valtonen90}, galaxy
mergers \citep{barnes+hernquist91}, group-cluster collisions \citep{bekki01},
 ram-pressure stripping \citep{gunn+gott72}, viscous
stripping \citep{nulsen82}, evaporation \citep{cowie+songalia77} and
`starvation' \citep{larson+80}. Although the effects,
time-scales and efficiencies of these physical processes can be
different, all together they serve to transform galaxies by disturbing
their kinematics, depleting their gas reservoir and so ultimately
quenching their star formation which can be even enhanced in the first
phase of the transformation.
Galaxies which pass through the cluster centre are those mostly affected by
such mechanism, while those just falling into the cluster outskirts or
moving on a tangential orbit may retain part of their gas. 
The impact of the large scale environment, such as the highest overdensities
(i.e. galaxy clusters and the so-called {\it superclusters}), on star
formation and other galaxy properties is however still
unclear. Superclusters provide and excellent chance to study such
effects. In these dynamically active and locally dense structures
processes such as cluster-cluster collisions and mergers in different
phases together with a wide and inter-connected range of environments
(from cluster cores to filaments) dramatically enhance the probability
to observe evidence of environmental effects on galaxy evolution at a
same epoch \citep[e.g.,][]{merluzzi+15}.

In this paper, we address the role of cluster mergers and group accretion in shaping the radio and optical properties of galaxies in the core region of the 
Shapley Concentration, where many clusters and groups in different evolutionary stages 
are located. In particular, we present a detailed radio-optical study of the BCGs and tailed radio 
galaxies of the A\,3558 and A\,3528 cluster complexes
based on data taken with the Giant Metrewave Radio Telescope (GMRT) and the Very Large Array (VLA), 
combined with the literature and archival information from the TIFR GMRT Sky Survey (TGSS), the Sydney University Molonglo Sky Survey (SUMSS), the VLA 
and the Australia Telescope Compact Array (ATCA). The global radio analysis spans almost two orders of magnitude in frequency, from 150 MHz to 8.46 GHz. Our analysis is complemented by the optical and near-IR information from the Shapley Supercluster Survey \citep[ShaSS,][]{merluzzi+15} together with the IR coverage with WISE.

The paper is organised as follows: 
in Sect. \ref{sec:shapley} we provide an overview of the region of the Shapley Concentration under investigation; 
in Sect. \ref{sec:obs} we describe the radio and optical observations and detail the radio analysis; 
the radio images and the galaxies sample are presented in Sects. \ref{sec:a3528} and \ref{sec:a3558}. 
The radio spectral study follows in Sect. \ref{sec:spx}. 
In Sect. \ref{sec:ottico} we present the results of the optical analysis. 
Finally, our results and conclusions are discussed and interpreted in Sects. \ref{sec:disc} and \ref{sec:conc}. 

Throughout the paper we use the convention $S_\nu\propto\nu^{-\alpha}$. We assume a standard cosmology with 
H$_0=70$ km s$^{-1}$ Mpc${-1}$, $\Omega_{\rm m}=0.3$ and $\Omega_\Lambda=0.7$ is assumed, which impies a 
conversion factor of 0.928 kpc/arcsec and an average luminosity distance 210 Mpc.


\begin{table*}[h!]
\caption[]{Logs of the observations}
\centering
\begin{tabular}{lccccccc}
\hline
\hline\noalign{\smallskip}
Cluster & RA, DEC (J2000) & Array & Project & Observing Date & $\nu$  & $\Delta\nu$ & Total time  \\
           &	                           &           &              &            & (MHz) &   (MHz)         & on source (h) 		  \\
\hline\noalign{\smallskip}
A\,3528~N & 12 54 20, --29 02 30 & GMRT    & 05TVa01 & 5/6-Apr-2004 &  235 &   8  &  4$^{\rm a}$  \\
                  & 12 54 20, --29 02 30 & GMRT    & 05TVa01  & 5/6-Apr-2004 &  610 &  16 &  4$^{\rm a}$   \\
                  & 12 54 22, --29 01 02 & VLA-CnB & AV246    & 13-Mar-2000 & 8460  & 100 & 0.25$^{\rm b}$ \\
                  & 12 54 21, --29 04 16 & VLA-CnB & AV246    & 13-Mar-2000 & 8460  & 100 & 0.25$^{\rm b}$ \\
A\,3528~S & 12 55 00, --29 40 00 & GMRT    & 05TVa01  &5/6-Apr-2004 &  235 &   8  &  4$^{\rm a}$  \\
                  & 12 55 00, --29 40 00 & GMRT    & 05TVa01  & 5/6-Apr-2004 &  610 &  16 &  4$^{\rm a}$  \\
                  & 12 54 51, --29 16 20 & VLA-CnB & AV246  & 3-Mar-2000 & 8460  & 100 & 0.25$^{\rm b}$ \\
A\,3532      & 12 56 30, --30 30 00 &  GMRT    & 05TVa01 & 5/6-Apr-2004 &  235 &   8  &  4$^{\rm a}$  \\
                  & 12 56 30, --30 30 00  & GMRT     & 05TVa01 & 5/6-Apr-2004 &  610 &  16 &  4$^{\rm a}$  \\ 
A\,3556      & 13 24 00, --31 38 00 &  GMRT    & 05TVa01  & 5/6-Apr-2004 &  235 &   8  &  4$^{\rm a}$  \\
                   & 13 24 00, --31 38 00 & GMRT     & 05TVa01  &5/6-Apr-2004 &  610 &  16 &  4$^{\rm a}$  \\ 
A\,3558       & 13 27 54, --31 29 32 & GMRT    & 22\_039  & 30-Aug-2012 &  325  &  32  &  8$^{\rm c}$  \\
                   & 13 27 54, --31 29 32 & GMRT    & 22\_039  & 02-May-2015 & 610 &  32 &  7$^{\rm c}$  \\
\hline\noalign{\smallskip}
\end{tabular}
\tablefoot{$^{\rm a}$ Observations of project 05TVa01 were carried out on two consecutive days with the
dual 235/610 MHz receiver, for a total of 8 hours each day. We cycled among the four pointing centres every
20 minutes. $^{\rm b}$ The total duration of project AV246 was one hour, and the observations were carried 
out switching among the three pointing centres every 10 minutes. $^{\rm c}$ Full track observations.}
\label{tab:logs}
\end{table*}
%

%

\section{The Shapley Concentration in context}\label{sec:shapley}
The Shapley Concentration \citep{shapley30} is the richest and most massive concentration of galaxy clusters, i.e. {\it supercluster}, in the local 
Universe \citep[e.g.,][]{raychaudhury89,scaramella+89,vettolani+90,zucca+93}. It is located in the Southern 
sky and lies behind the Hydra-Centaurus cluster. Overall, the structure covers a redshift range $0.033 \simlt z \simlt 0.06$
\citep{quintana+95,quintana+97}, 
with a mean redshift $z\approx0.048$. Due to the very high overdensity of 
galaxy clusters, the innermost region of the Shapley Concentration is dynamically active, and thanks to its proximity it is 
an ideal place to investigate the effects of  group accretion and cluster mergers.
The masses and the bolometric luminosities of the individual clusters and
groups range between 
$M_{\rm 500}\approx0.1-6\times10^{14}$ h$^{-1}$ M$_{\odot}$\footnote{M$_{\rm 500}$ is defined as the mass calculated at the radius where the cluster density exceeds 500 times the critical density of the Universe.} \citep{reisenegger+00} and 
$L_{\rm X} \approx0.5-6.7\times10^{44}$ erg s$^{-1}$ \citep{defilippis+05}, respectively.

The A\,3558 cluster complex is considered the centre of the Shapley Concentration Core and consists of a chain of three 
Abell clusters (A\,3556, A\,3558 and A\,3562) and two smaller groups (SC\,1327--312 and SC\,1329--313).
It extends for about 7.5 h$^{-1}$ Mpc in the East-West direction, at an average redshift of $\langle z \rangle\approx0.048$.
Studies in the optical \citep{bardelli+98a,bardelli+98b,merluzzi+15}, X--ray \citep{markevitch+vikhlinin97,ettori+00,rossetti+07,ghizzardi+10}, and radio \citep{venturi+00,venturi+03,giacintucci+04,giacintucci+05,venturi+17} bands provide several pieces of evidence in
support of the idea that the whole region between A\,3558 and A\,3562 is unrelaxed.

The A\,3528 cluster complex is located at a projected distance of approximately 19 h$^{-1}$ Mpc North-West of A\,3558,
and it is formed by the three Abell clusters A\,3528, A\,3530 and A\,3532. 
It extends for about 7.5 h$^{-1}$ Mpc$^{-1}$ in the North-South direction, at the average redshift  $\langle z \rangle\approx0.054$. 
The X-ray emission of A\,3528 itself \citep{schindler+96,gastaldello+03} is in the form of two sub-clumps, 
termed A\,3528~N and A\,3528~S, centred on the two dominant galaxies. 
Despite that, the X-ray analysis showed that the two sub-clumps, and also A\,3530 and A\,3532,
have an overall relaxed appearance in the X--ray, and are classified as cool-cores 
 \citep{gastaldello+03,lakhchaura+13}.
\cite{lakhchaura+13} suggested the presence of cavities around A\,3532, typically found in relaxed clusters. 
\medskip


\begin{figure*}[h!]
\centering
\includegraphics[width=0.9\textwidth]{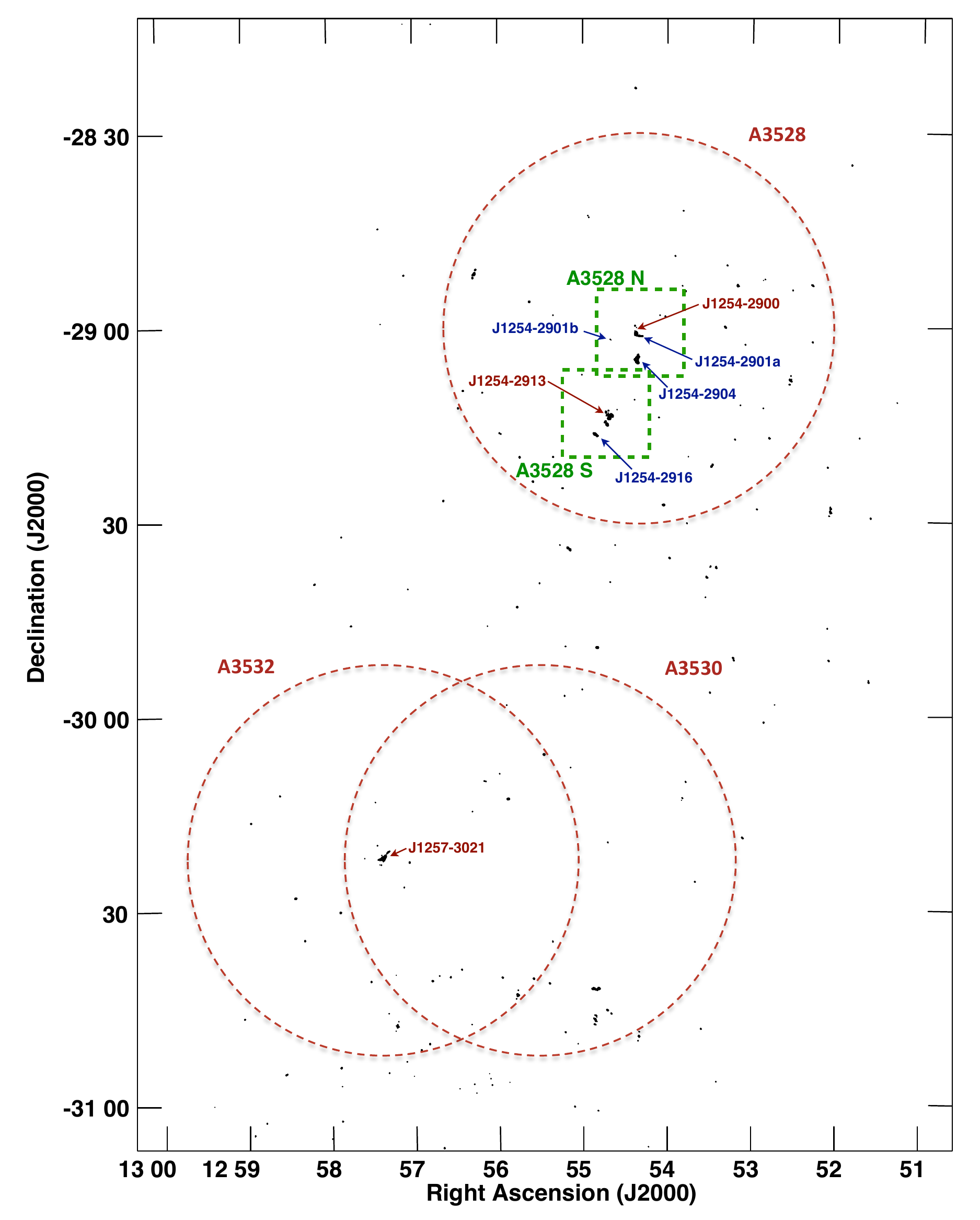}
\caption{Radio contours of the A\,3528 complex at 610 MHz. The restoring beam is $6.8^{\prime\prime}\times4.7^{\prime\prime}$, p.a. $29^\circ$ and the noise level of 0.16 mJy beam$^{-1}$. The levels are $7\sigma\times(1, 2, 4, 8, 16, 32, 128)$. The dashed red circles show the Abell radius of the three clusters. The BCGs are labelled in red, the tailed radio galaxies in blue. The dashed green squares represent the A\,3528~N and A\,3528~S sub-clusters, based on the X-ray observations \citep{gastaldello+03}.}
\label{fig:mosaic_A3528}
\end{figure*}


\begin{figure*}[h!]
\centering
\includegraphics[width=0.85\textwidth]{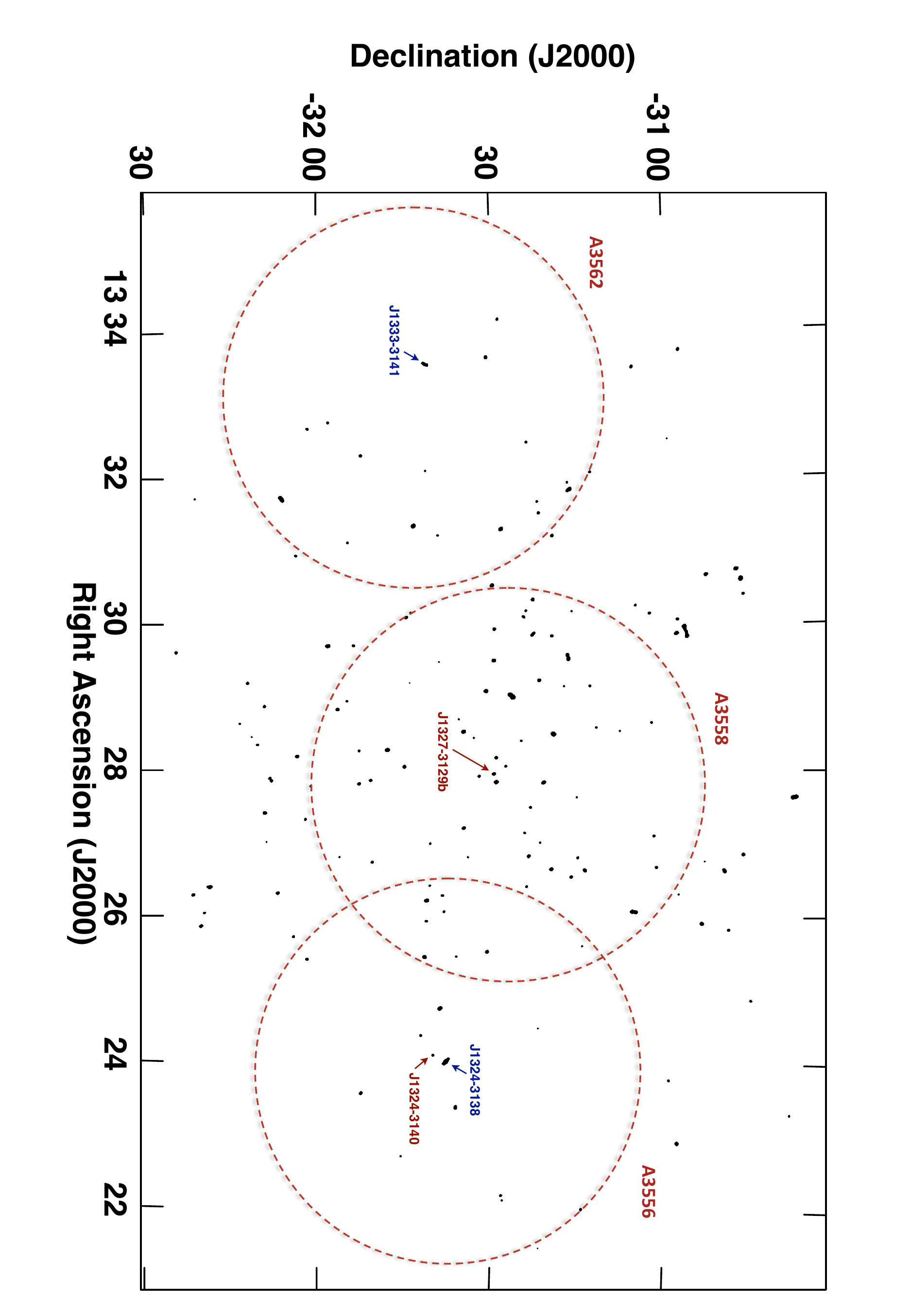}
\caption{Radio contours of the A\,3558 complex at 325 MHz. The restoring beam is $14.05^{\prime\prime}\times9.53^{\prime\prime}$, p.a. equal to $18.39^\circ$ and the noise level of  0.09 mJy beam$^{-1}$. The levels are $7\sigma\times(1, 2, 4, 8, 16, 32, 128)$. The dashed red circles show the Abell radius of the three clusters. The BCGs are labelled in red, the tailed radio galaxies in blue.}\label{fig:mosaic_A3558}
\end{figure*}

%

\section{Observations and data reduction}\label{sec:obs}
\subsection{Radio data}

We observed the A\,3528 and A\,3558 complexes with the GMRT and with the VLA to image (i) the radio emission 
from the BCGs in all clusters, classify their morphology and study their spectral properties, and (ii) other 
extended cluster radio galaxies, derive their spectral properties and possible connection with the cluster merger. 
To reach our goals, we performed observations over a wide range of frequencies and angular resolutions. 
Table \ref{tab:logs} provides the details of the observations.

\subsubsection{The GMRT observations}\label{sec:gmrt}
The A\,3528 complex and the cluster A\,3556 were observed on 5th and 6th April 2004, with the dual band configuration 
at 235/610 MHz. The set up used the RR polarisation at 610 MHz and the LL polarisation at 235 MHz, with bandwidths of 16 and 
8 MHz respectively.  A\,3558 was observed in a separate project at 325 MHz and at 610 MHz, with 
$\Delta\nu=32$ MHz. 
Both polarisations were recorded at the two frequencies.

The data were collected in spectral-line mode at all frequencies, that is 128 and 64 channels at 610 and 235 MHz for
project 05TVa01, and 256 channels at 325 MHz and 610 MHz for project 22\_039.  
The calibration and data reduction were performed using the NRAO Astronomical Image Processing System (AIPS)\footnote{http://aips.nrao.edu/} package. The raw data were first processed with the software {\it flagcal} \citep{prasad+12} to
remove RFI and apply bandpass calibration. Then further editing, self-calibration and imaging were performed. To find a
compromise between the size of the dataset and the need to minimise bandwidth smearing effects within the primary beam, 
after bandpass calibration the channels in each individual dataset of project 05TVa01 were averaged to 6 channels of 
$\approx1$ MHz each at 235 MHz, and 
$\approx2$ MHz each at 610 MH, while for project 22\_039 they were averaged to 25 and 50 channels of $\approx1$ and 
0.5 MHz each at 610 MHz and 325 MHz respectively.
The {\it a-priori} calibration was performed using 3C286 and 1311--222 (3C283) as primary and secondary calibrators, respectively.
The Baars \citep{baars+77} flux density scale was used.
The dual band observations at 235/610 MHz (project 05TVa01)  were split into four pointings, centred on the three 
individual clusters  of the A\,3528 complex and on A\,3556 (see Table \ref{tab:logs}), while one single pointing on A\,3558 was used 
for project 22\_039. The primary beam of each pointing is about $0.9$, $1.8$ and $2.5$ deg at 610, 325 and 235 MHz
respectively. Hence, at all frequencies we used a multi-facet clean to take into account the sky curvature. 
In addition, direction dependent calibration was used for project 22\_039 using the task \texttt{PEELR} in AIPS.

After a number of phase-only and one phase and amplitude self-calibration cycles, we produced the final images at each 
frequency for each day of observation.  The three datasets on the A\,3528 complex at 235 MHz and 610 MHz were 
self-calibrated and primary-beam corrected separately and then combined to make a final mosaic. 
At each frequency we produced final images at full (weighting Briggs and robust 0) and at lower resolutions (weighting Briggs and robust 1, with
different uv tapers depending on the frequency, and using a maximum of 60\% of the full uv-coverage) in search for further extension of the radio galaxies.
We did not
detect further emission at low resolution at any frequency
(at the sensitivity level of our images), hence the analysis and discussion in this paper
is fully based on the full resolution images, which also have the best quality.
The average residual amplitude errors in our data are $\leq5\%$ at 610 and 325 MHz, and about 8\% at 235 MHz. 
In Table  \ref{tab:array_res} we provide the relevant
information on the final images for each cluster, i.e., angular resolution and 1$\sigma$ noise level. 

Figures \ref{fig:mosaic_A3528} and \ref{fig:mosaic_A3558} provide an overview of the radio emission of the A\,3258 complex (at 610 MHz) and of the A\,3558 complex (at 325 MHz) respectively.

The analysis and discussion in this paper
is fully based on the full resolution images, which we present here, since they recover the whole extent of the radio emission 
in all sources and provide the highest quality.


\begin{table}[h!]
\caption[]{Observational parameters of the final images.}
\centering
\resizebox{0.47\textwidth}{!}{%
\begin{tabular}{lccccc}
\hline
\hline\noalign{\smallskip}
Cluster & Array &  $\nu$ & Resolution & P.A. & rms \\ 
            &           &   (MHz) & ($^{\prime\prime}\times^{\prime\prime}$) & ($^{\circ}$)  & (mJy~beam$^{-1}$) \\
\hline\noalign{\smallskip}

A\,35288~N & GMRT  &   235  &  15.7 $\times$11.8 & 4.3 & $\sim$ 0.50 \\				
                 &             &  610   &  6.2$\times$4.3       & 24.4  & $\sim$ 0.10  \\
                 &  VLA-CnB    &  8460  & 2.7$\times$1.5     &  52.8  & $\sim$0.015  \\                
A\,3528~S & GMRT  &   235  &  15.7 $\times$11.8 & 4.3 & $\sim$0.50 \\				
                 &             &  610   &  6.2$\times$4.3       & 24.4  & $\sim$0.10  \\   
                 &  VLA -CnB   &  8460  &  2.9$\times$1.4   &  52.4  & $\sim$0.015 \\            
A\,3532 & GMRT  &   235  &  15.7 $\times$11.8 & 4.3 & $\sim$0.50 \\	
                 &          &  610   &  6.2$\times$4.3        & 24.4  & $\sim $0.13  \\			
A\,3556 & GMRT  &   235  &  16.0 $\times$11.5 & 13.7 & $\sim$0.65 \\				
                &           &  325  &  14.1$\times$9.5    & 18.4  & $\sim$0.08   \\                 
                &          &  610   & 7.0$\times$4.4  &    31.3   & $\sim$0.09   \\
A\,3558    & GMRT  &  325  &  14.1$\times$9.5    & 18.4  & $\sim0.12$   \\               
                &           &  610   &  9.6$\times$5.0   &  32.8  & $\sim$0.06  \\
\hline\noalign{\smallskip}
\end{tabular}}
\tablefoot{The noises in the last column are an average of the whole map; ``local'' noises are calculated for the sigle sources and shown in the label of each figure.}
\label{tab:array_res}
\end{table}


\subsubsection{The VLA observations}\label{sec:vla}
A\,3528~N and A\,3528~S  were observed with the VLA in the hybrid CnB configuration at 8.46 GHz
with 2 IFs and a total bandwidth $\Delta\nu=100$ MHz (see Table \ref{tab:logs}), as part of an earlier
separate project. Part of those observations were published in \cite{venturi+03}.
Following a standard approach, the data were edited, self-calibrated and imaged using AIPS. 
The 1$\sigma$ rms is $\approx 0.015$ mJy~beam$^{-1}$, while the residual amplitude errors are 
$\sim$3\% (Baars flux density scale). The most relevant parameters of 
the final full resolution image are reported in Table \ref{tab:array_res}.

\subsubsection{Complementary data}\label{sec:complementary_info}
In order to complete the spectral coverage of our analysis we added the information from literature and archival data from  
the TIFR GMRT Sky Survey (TGSS) at 150 MHz, the Sydney University Molonglo Sky Survey (SUMSS) Source
Catalog at 843 MHz, the NRAO VLA Sky Survey (NVSS) at 1.4 GHz and the Australia Telescope Compact array (ATCA) at 
1.38, 2.38 and 4.79 GHz. In Table \ref{tab:archive} we report the resolution and the noise of the images we used and
refer to the original papers.
%

\begin{table}[h!]
\caption[]{Observational parameters of the literature images. The information refers to the A\,3528 and A\,3558 cluster complexes, on the top and on the bottom, respectively.}
\centering
\resizebox{0.47\textwidth}{!}{%
\begin{tabular}{ccccc}
\hline
\hline\noalign{\smallskip}
Array &  $\nu$ & Resolution & P.A. & rms \\ 
   &   (MHz) & ($^{\prime\prime}\times^{\prime\prime}$) & ($^{\circ}$)  & (mJy~beam$^{-1}$) \\
\hline\noalign{\smallskip}
TGSS - GMRT$^{\rm a}$    &   150  &  24.0$\times$15.0 & 30 &  9 \\						
ATCA$^{\rm b}$                  &  1380  & 10.0$\times$6.0  &  0 & 0.14 \\				     	
ATCA$^{\rm c}$                  &  1400  & 11.8$\times$6.2  &  -0.31  & 0.56 \\				
NVSS - VLA-D$^{\rm d}$    &  1400  & 45.0$\times$45.0  & 61  & 0.13 \\ 
ATCA$^{\rm b}$     &  2380  &  6.0$\times$3.5  &  2  & 0.15 \\					        
ATCA$^{\rm c}$     &  2400  &  6.2$\times$4.2  &  0.15  & 0.39 \\			         	
\hline\noalign{\smallskip}
VLA$^{\rm g}$ 	    & 327       &   $59.0\times45.0$ & 80.5 &1.9 \\				
SUMSS - MOST$^{\rm e}$    &  843  & 82.0$\times$43.0  &  0  &  1.4 \\					
ATCA$^{\rm f}$     &  1380  & 6.0$\times$10.0  & 0 & 0.2 \\					
ATCA$^{\rm g}$    &  1376  & 10.2$\times$6.0  &  0.26 & 0.16 \\			 	
ATCA$^{\rm f}$     &  2380  &  5.3$\times$3.4  & 0 & 0.15 \\					
ATCA$^{\rm g}$    &  4790  & 20.2$\times$10.0  &  0  &  0.04 \\				        
VLA$^{\rm g}$      & 8640 & $20.0\times10.0$ & 0 & 0.04 \\					
\hline\noalign{\smallskip}
\end{tabular}}
{\bf References.} $\rm ^a~$TGSS image (http://tgss.ncra.tifr.res.in); $\rm ^b~$\cite{venturi+01}; $\rm ^c~$\cite{reid+98}; $\rm ^d~$NVSS image (http://www.cv.nrao.edu/nvss/); $\rm ^e~$SUMSS catalog (http://www.physics.usyd.edu.au/sifa/Main/SUMSS); $\rm ^f~$\cite{venturi+97}; $\rm ^g~$\cite{venturi+98}.
\label{tab:archive}
\end{table}

\subsection{New optical-NIR data and archive data} 
A multi-wavelength survey of the whole Shapley supercluster 
(Shapley Supercluster Survey, ShaSS) was performed by \cite{merluzzi+15} with the aims to
investigate the role of the mass assembly on galaxy evolution. The
survey covers 23\,deg$^2$ (i.e. 260 Mpc$^2$ at cluster redshift) centered on A\,3558 and combines observations in the ESO-VST $ugri$ filters
with ESO-VISTA $K$-band imaging. In the framework of this project,
ShaSS has recently covered the A\,3528 complex (12\,deg$^2$ in
$i$ band and 3\,deg$^2$ in $g$ band) providing new deep photometry for
all the galaxies presented in this work. At the distance of the Shapley
Concentration, the ShaSS imaging has a spatial resolution $\lesssim 1$\,kpc
which allows us to study the structure of the individual galaxies.

Furthermore we made use of the
imaging with the Wide-Field Infrared Survey
Explorer \citep[WISE; ][]{wright+10} in the three channels W1-W3
spanning the wavelength range 3.4-12\,$\mu$m.

\begin{figure*}[h!]
\centering
{\includegraphics[width=0.45\textwidth]{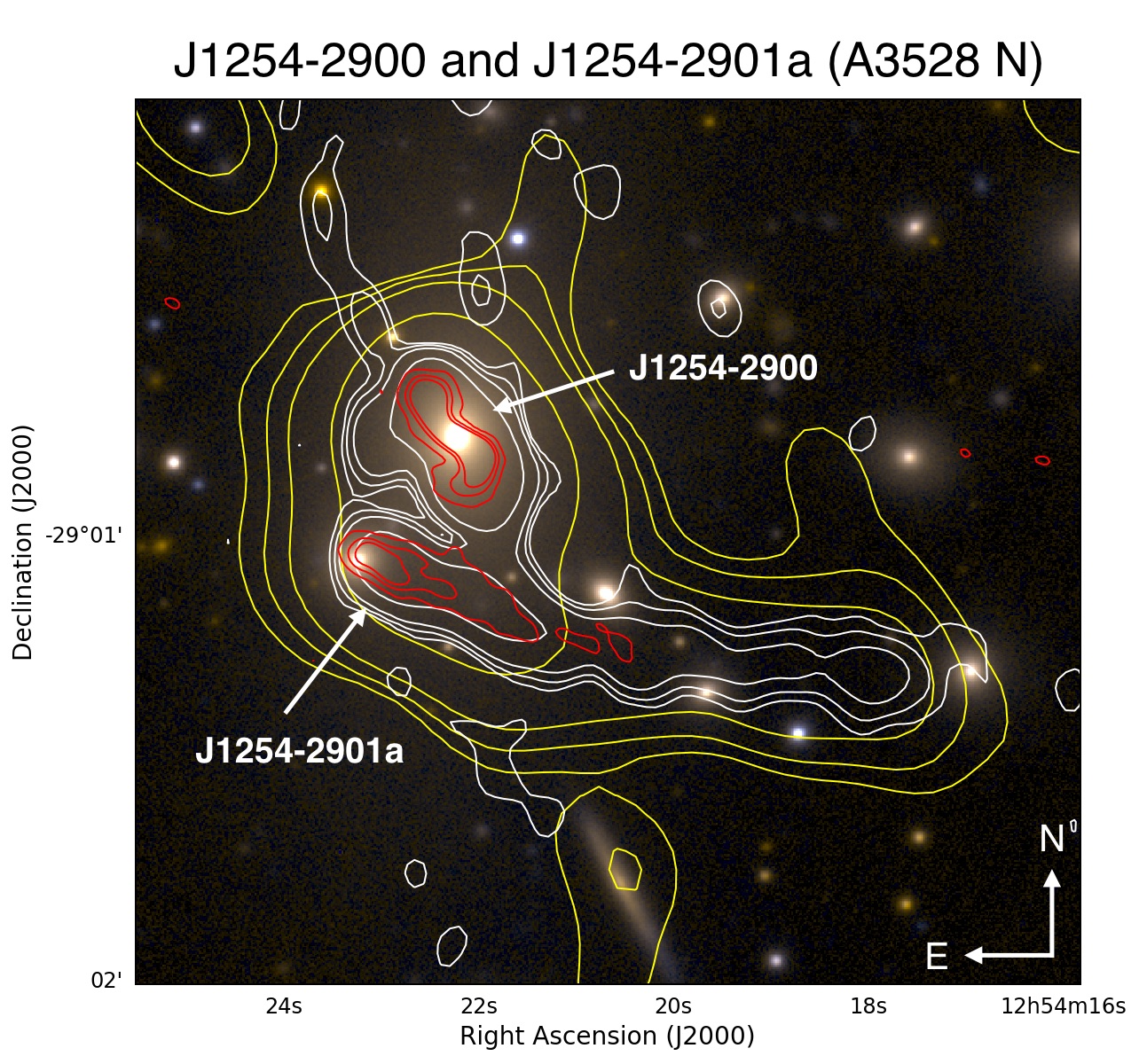}}
\hspace{2mm}
{\includegraphics[width=0.45\textwidth]{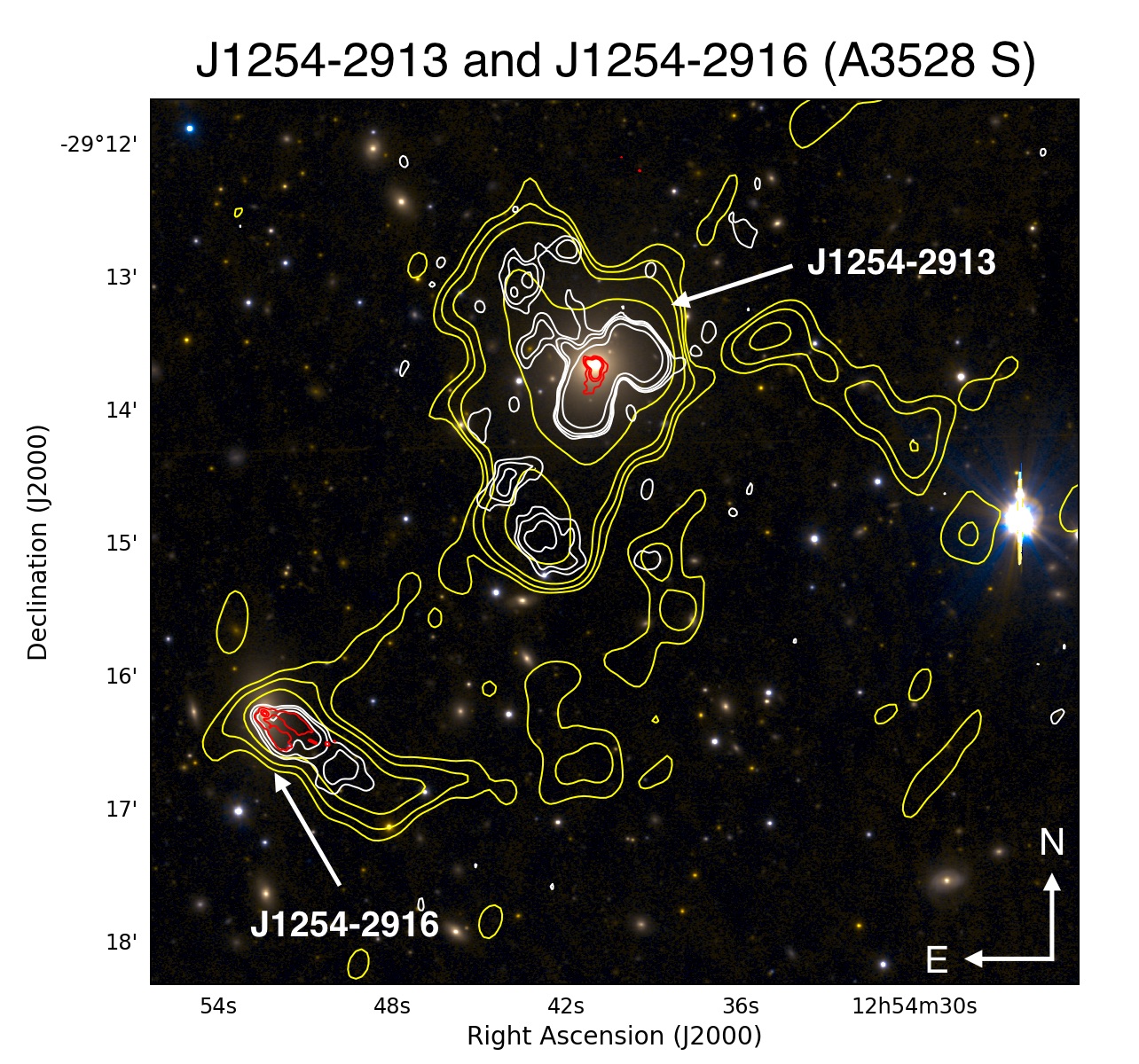}}
\hspace{2mm}
{\includegraphics[width=0.45\textwidth]{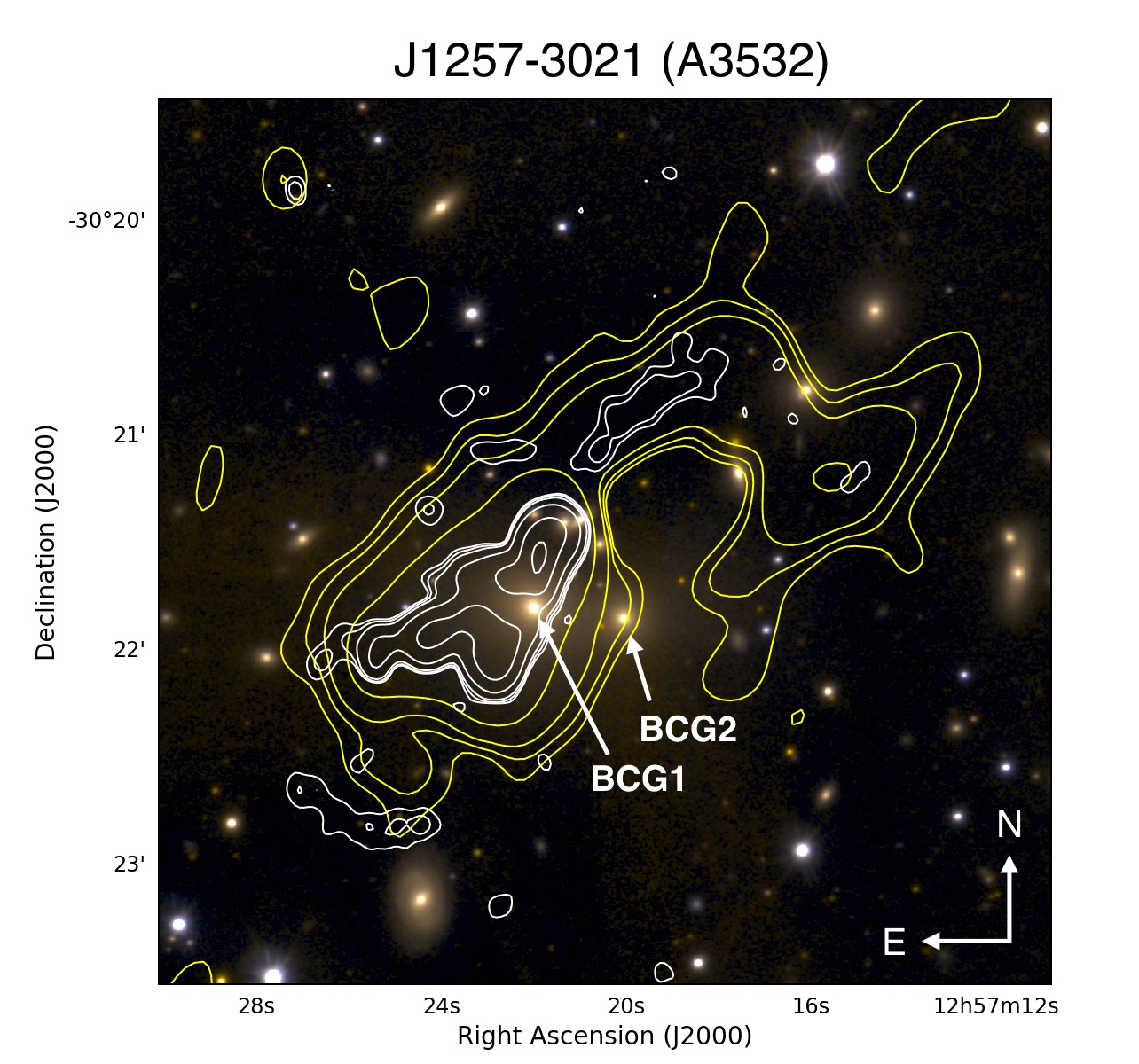}}
\caption{VST-ACESS {\it gri} composite images with GMRT radio contours at 235 (yellow) and 610 (white) MHz and VLA radio contours at 8.4 GHz (red). {\it Top left panel:} J\,1254-2900 and J\,1254-2901a (A\,3528~N). {\it Top right panel:}  J\,1254-2913 and J\,1254-2916 (A\,3528~S). The 235 MHz contours are at $3\sigma\times(1, 2, 4, 8, 16)$ mJy beam$^{-1}$, the resolution is $16.3^{\prime\prime}\times11.9^{\prime\prime}$, p.a. $9.5^\circ$ and noise level $\sigma_{\rm 235~MHz}=1.26$ mJy beam$^{-1}$. The 610 MHz contours are at $3\sigma\times(1, 2, 4, 8, 16)$ mJy beam$^{-1}$, the resolution is $6.8^{\prime\prime}\times4.7^{\prime\prime}$, p.a. $29^\circ$ and noise level $\sigma_{\rm 610~MHz}=0.21$ mJy beam$^{-1}$. The 8.4 GHz contours are at $3\sigma\times(1, 4, 8)$ mJy beam$^{-1}$, the resolution is $2.76^{\prime\prime}\times1.59^{\prime\prime}$, p.a. $49.82^\circ$ and noise level $\sigma_{\rm 8.4~GHz}=0.05$ mJy beam$^{-1}$ . {\it Bottom panel:} J\,1257-3021 (A\,3532). The 235 MHz contours are at $3\sigma\times(1, 2, 4, 8, 16)$ mJy beam$^{-1}$, the resolution is $16.3^{\prime\prime}\times11.9^{\prime\prime}$, p.a. $9.5^\circ$ and noise level $\sigma_{\rm 235~MHz}=0.76$ mJy beam$^{-1}$. The 610 MHz contours are at $3\sigma\times(1, 2, 4, 8, 16, ...)$ mJy beam$^{-1}$, the resolution is $6.8^{\prime\prime}\times4.7^{\prime\prime}$, p.a. $29^\circ$ and noise level $\sigma_{\rm 610~MHz}=0.14$ mJy beam$^{-1}$.}
\label{fig:opt-radio_bcgA3528}
\end{figure*}

\section{The radio galaxy population in the A\,3528 Complex}\label{sec:a3528}
A statistical study of the radio galaxy population in the A\,3528 complex was performed using 
ATCA and was published in  \cite{venturi+01}.
Here we present a detailed morphological analysis of the radio emission from the BCGs and of the galaxies showing extended
radio tails, while we present the spectral analysis in Sect. \ref{sec:spx}. 

The sample includes the three BCGs J\,1254-2900 (in A\,3528~N), J\,1254-2913 (in A\,3528~S) and J\,1257-3021 (A\,3532) 
and four tailed radio galaxies: J\,1254-2901a, J\,1254-2901b and J\,1254-2904 in A\,3528~N, and J\,1254-2916 in A\,3528~S.  
The properties of the galaxies are reported in the upper part
of  Table \ref{tab:cluster}. The dominant galaxy of the cluster A\,3530 is radio quiet, and it is reported in Table \ref{tab:cluster} for completeness.


\subsection{Radio properties of the BCGs}\label{sec:bcg_a3528}
\subsubsection{J\,1254-2900 (A\,3528~N)}\label{sec:j2900}
The radio emission from the BCG in A\,3528~N at 1.4 and 2.3 GHz was first reported in \cite{reid+98} and \cite{venturi+01}, who described it as a small double radio galaxy (largest projected linear size $\sim$ 30 kpc), fully contained
within the faint optical halo emission. This is confirmed also at the lower frequencies presented in this paper, where no
further extension is detected either at 610 MHz or at 235 MHz. 
 At 8.4 GHz the radio emission has a remarkable S shape. The origin of
S- and X-shaped radio galaxies is still debated, and one of the
possibilities is the merger between two galaxies each hosting a super
massive black hole \citep{gopal-krishna+12}. Moreover, reorientation of the radio jets has been proposed for a number of radio
galaxies \citep{liu+18}.
Table \ref{tab:cluster} reports the radio power at 610 MHz, which is typical of the 
FRI/FRII transition value. The top left panel  of Fig. \ref{fig:opt-radio_bcgA3528} shows the radio emission associated with the 
BCG and with 
the head-tail source J\,1254-2901a (see Sect. \ref{sec:j2901a}). At 235 MHz the radio emission of the BCG blends with that of
the head-tail.

\subsubsection{J\,1254-2913 (A\,3528~S)}\label{sec:j2913}
The radio emission associated with the BCG in A\,3528~S is shown in the top right  panel  of Fig. \ref{fig:opt-radio_bcgA3528}, together with the head-tail J\,1254-2916 (see Sect. \ref{sec:j2916}). J\,1254-2913 is quite complex and difficult to classify. At 8.4 GHz the emission  is resolved in three components, i.e. a ``core'' and two additional features (see also left panel in Fig.\ref{fig:J1254-2913_morpho}), and it is fully
contained in the most central region of the host galaxy (see also Sect. \ref{sec:ottico}).
The emission at 610 MHz extends on a scale slightly larger
than that of the optical image, with a morphology which could be classified as wide-angle tail. At this frequency the
radio emission is edge brightened, as clear from the top right panel in Fig. \ref{fig:opt-radio_bcgA3528}, and its radio power is the highest in our sample (see Table 
\ref{tab:cluster}). Some residual positive emission 
is visible at some distance from the BCG, suggesting the presence of more extended emission unrecoverable with
the current 610 MHz data. At 235 MHz the radio emission encompasses the residuals detected at 610 MHz 
North-East and South of the BCG, and further hints of residuals are detected in the form of filaments
West of the BCG. Such residuals of very low
surface brightness are clearly separated from the brightest part of the emission. They extend $\sim$ 100 kpc and are suggestive of further very low surface brightness extended emission, whose
origin is under investigation and will be presented in a future work.

\subsubsection{J\,1257-3021 (A\,3532)}\label{sec:j3021}
The dominant galaxy in A\,3532 is a well-known dumb-bell, i.e. two galaxies of roughly equal brightness inside a common halo,
whose radio emission has been studied at GHz frequencies
\citep{gregorini+94} and at 610 MHz \citep{lakhchaura+13}. Only the easternmost nucleus (i.e. BCG1, see bottom panel in Fig. \ref{fig:opt-radio_bcgA3528})
is radio loud, with the second highest radio power among our sample (see Tab. \ref{tab:cluster}). Radio contours from our observations
are given in the bottom panel of Fig. \ref{fig:opt-radio_bcgA3528}. At GHz frequencies and at 610 MHz the radio galaxy has a double morphology,
with tails which bend asymmetrically to the East, suggesting a motion towards the cluster A\,3530. As for J\,1254-2913 in A\,3528~S,
the radio emission is edge brightened. At 235 MHz the radio emission extends well beyond the optical region, in a
tail elongated in the North-West direction. Such tail shows a sharp 90$^{\circ}$ bend, after which the radio emission
broadens in a direction almost perpendicular to the tail itself. Unfortunately, no 8.4 GHz VLA observations are available for this BCG, while the literature observations at 5 GHz \citep{gregorini+94} do not show any morphological difference from our 610 MHz image, and for this reason these radio contours are not shown in the bottom panel of Fig. \ref{fig:opt-radio_bcgA3528}.

\begin{figure*}[h!]
\centering
{\includegraphics[width=0.45\textwidth]{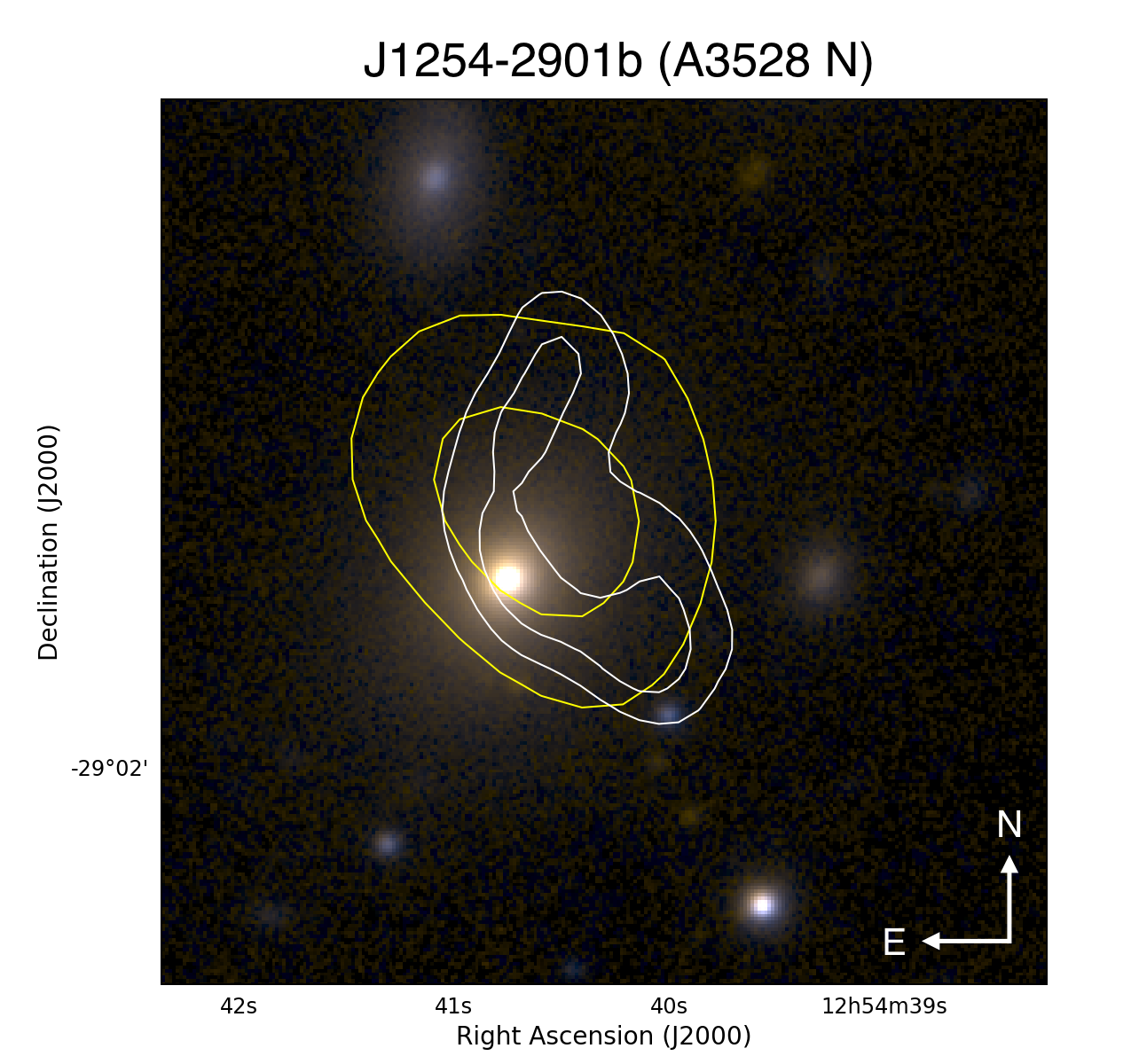}}
\hspace{2mm}
{\includegraphics[width=0.45\textwidth]{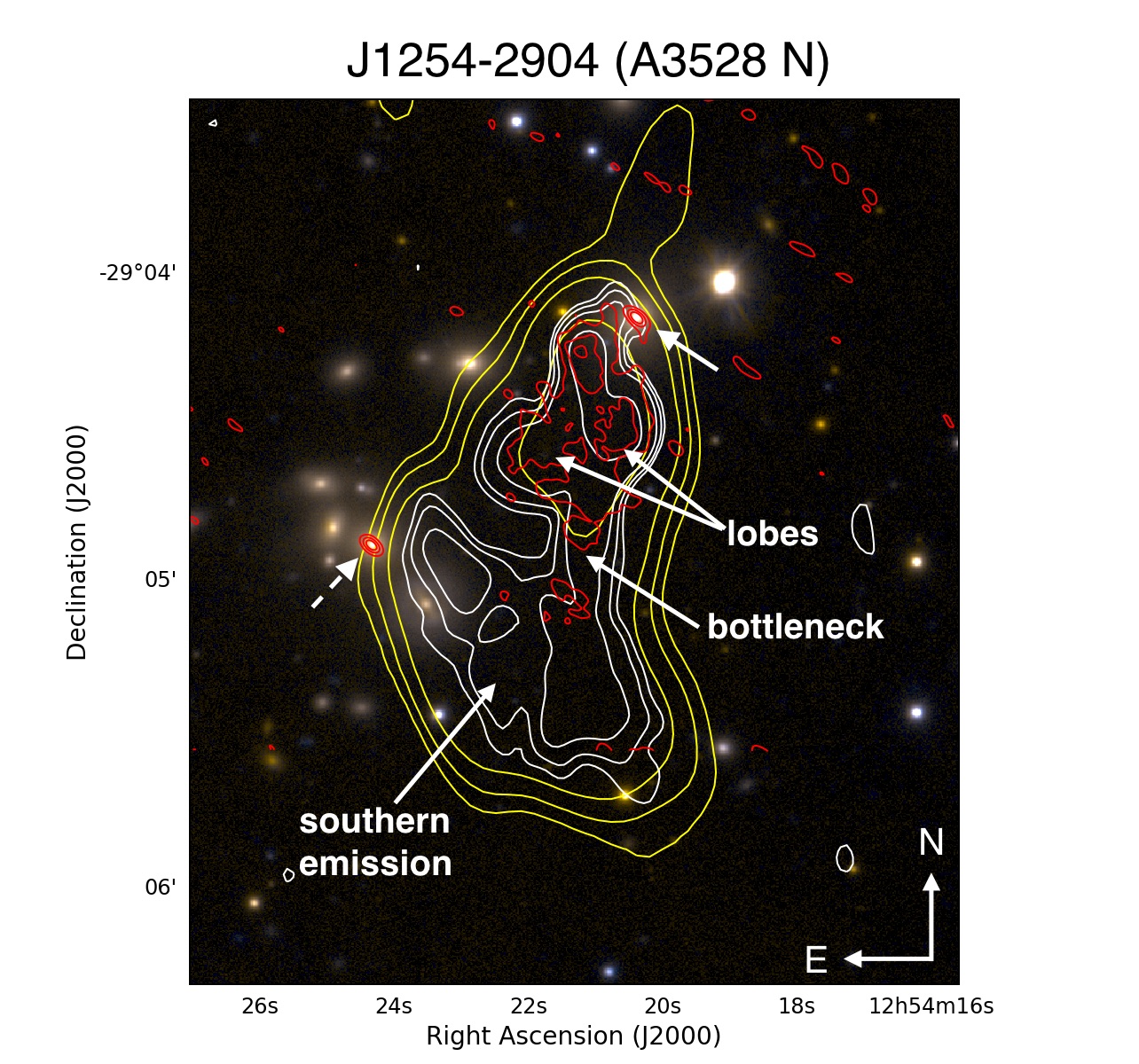}}
\hspace{2mm}
{\includegraphics[width=0.45\textwidth]{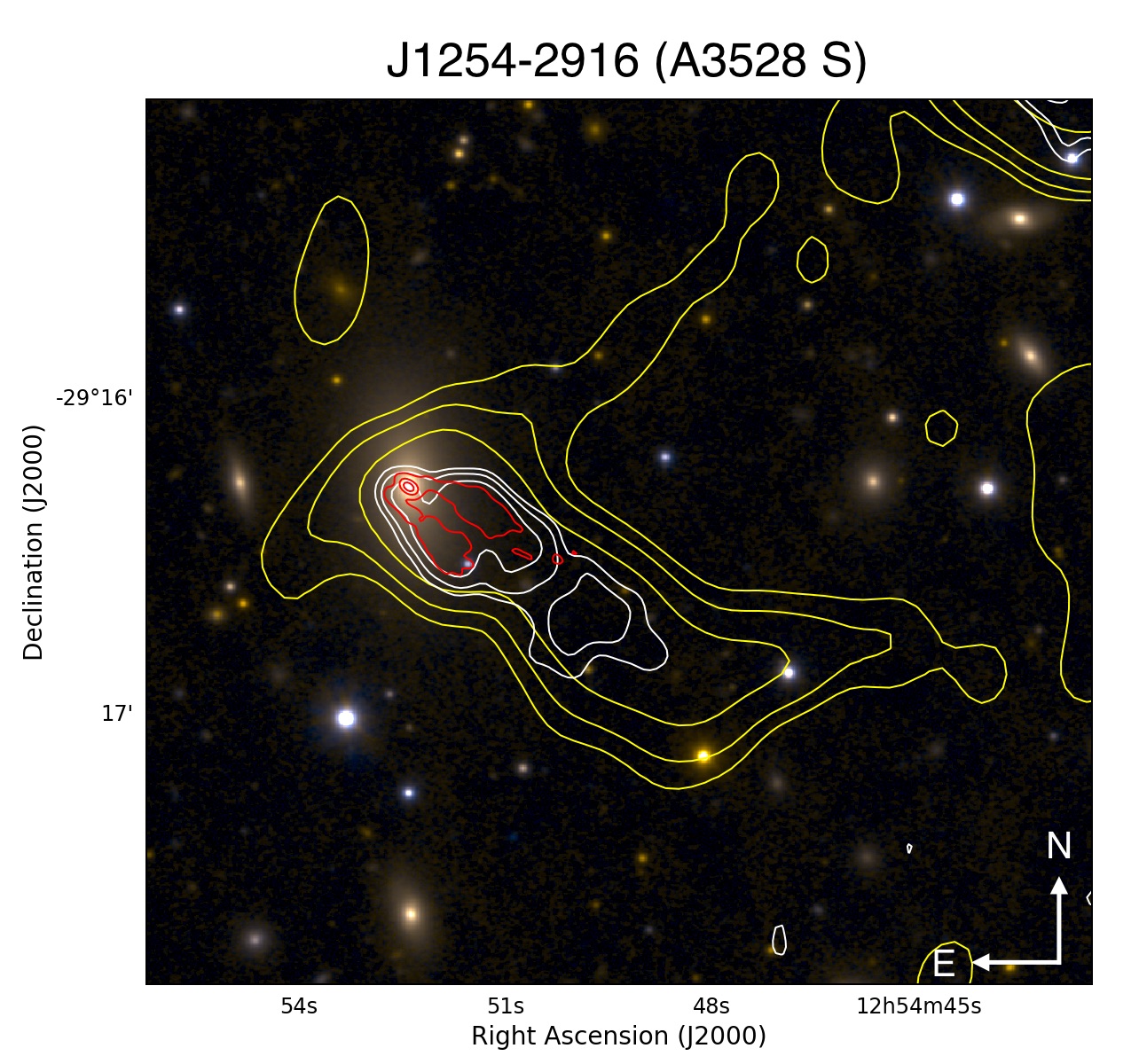}}
\caption{VST-ACESS {\it gri} composite images with GMRT radio contours at 235 (yellow) and 610 (white) MHz and VLA radio contours at 8.4 GHz (red). {\it Top left panel:} J\,1254-2901b (A\,3528~N). {\it Top right panel:} J\,1254-2904 (A\,3528~N). {\it Bottom panel:} J\,1254-2916 (A\,3528~S). The 235 MHz contours are at $3\sigma\times(1, 2, 4, 8, 16)$ mJy beam$^{-1}$, the resolution is $16.3^{\prime\prime}\times11.9^{\prime\prime}$, p.a. $9.5^\circ$ and noise level $\sigma_{\rm 235~MHz}=1.26$ mJy beam$^{-1}$. The 610 MHz contours are at $3\sigma\times(1, 2, 4, 8, 16)$ mJy beam$^{-1}$, the resolution is $6.8^{\prime\prime}\times4.7^{\prime\prime}$, p.a. $29^\circ$ and noise level $\sigma_{\rm 610~MHz}=0.21$ mJy beam$^{-1}$. The 8.4 GHz contours are at $3\sigma\times(1, 4, 8)$ mJy beam$^{-1}$, the resolution is $2.76^{\prime\prime}\times1.59^{\prime\prime}$, p.a. $49.82^\circ$ and noise level $\sigma_{\rm 8.4~GHz}=0.05$ mJy beam$^{-1}$.}
\label{fig:opt-radio_tailA3528}
\end{figure*}

 \begin{table*}[h!]
\caption[]{Summary of the properties of the galaxies in our sample. A\,3528 complex in the upper part, A\,3558 complex
in the lower part.}
\centering
\begin{tabular}{llcccccc}
\hline
\hline\noalign{\smallskip}
Cluster & Galaxy & RA$_{\rm J2000}$ & DEC$_{\rm J2000}$ & $z$ & optical  & radio & logP$_{\rm 610~MHz}$\\
name & name & ($\rm^h$~$\rm^m$~$\rm^s$)  & ($^\circ$~$^\prime$~$^{\prime\prime}$) & & morphology & 
morphology & W~Hz$^{-1}$\\
\hline\noalign{\smallskip}
A\,3528~N & J\,1254-2900     & 12~54~22.1 & --29~00~48 &  0.0541 & cD (BCG) & D & 24.47 \\
A\,3528~N & J\,1254-2901a   & 12~54~22.9 & --29~01~02 & 0.0544& ellipt. & HT & 24.12 \\
A\,3528~N & J\,1254-2904     & 12~54~20.4 & --29~04~09 & 0.0545 & ellipt. & NAT & 24.41 \\
A\,3528~N  & J\,1254-2901b  & 12~54~40.7 & --29~01~49 & 0.0529 & ellipt. & NAT & 22.94 \\
A\,3528~S & J\,1254-2913 & 12~54~41.0 & --29~13~39 & 0.0573 & cD (BCG) & WAT & 25.23 \\
A\,3528~S & J\,1254-2916 & 12~54~52.4 & --29~16~18 &  0.0481 & ellipt. & HT & 23.68 \\
A\,3530 & J\,1255-3019 & 12~55~34.5 & --30~19~50  & 0.0537 &   ellipt. & undet. & --\\     
A\,3532 & J\,1257-3021 & 12~57~22.5 & --30~21~45 & 0.0541 &  dumb-bell (BCG) & WAT & 25.03 \\
\hline\noalign{\smallskip}
A\,3556 & J\,1324-3138 & 13~23~57.5 & --31~38~45 & 0.0502 & ellipt. & HT & 23.89 \\
A\,3556 & J\,1324-3140 & 13~24~06.7 & --31~40~12 & 0.0480 & cD (BCG) & D & 23.01 \\
A\,3558 & J\,1327-3129b & 13~27~56.8 & --31~29~43 & 0.0469 & cD (BCG) & unres. & 22.88 \\
A\,3562 & J\,1333-3140  & 13~33~34.8 & --31~40~21 & 0.0488 & cD (BCG) & undet. & -- \\
A\,3562 & J\,1333-3141 & 13~33~31.6 & --31~41~01 & 0.0501 & ellipt. & HT & $23.58^\star$ \\
\hline\noalign{\smallskip}
\end{tabular}
\tablefoot{D=double; HT=head tail; NAT=narrow-angle tail; WAT=wide-angle tail; cD= central dominant. $^\star$ this value was obtained from \cite{venturi+03}.}
\label{tab:cluster}
\end{table*}

\subsection{Tailed radio galaxies}\label{sec:tail_a3528}
\subsubsection{J1254-2901a (A\,3528~N)}\label{sec:j2901a}
This source is located immediately South of the BCG in A\,3528~N (e.g. J1254-2900, see the Sect. \ref{sec:j2900}) and it is clearly a head-tail radio 
galaxy. The length of the tail depends on the observing frequency: it is approximately 80$^{\prime\prime}$ at 235 MHz and 
610 MHz ($\sim$ 80 kpc at source redshift), and it is considerably shorter at higher frequencies, suggesting a spectral 
steepening away from the nucleus. As clear from Fig. \ref{fig:opt-radio_bcgA3528} (top left panel), the head of the tail remains 
unresolved even at 8.4 GHz, where we barely see the bifurcation of the inner part of the jets. The tail is not straight: the inner
$\sim40^{\prime\prime}$ are slightly oriented towards South, while the remaining part of the tail is aligned in the
East-West direction. Overall, the tail suggests that the projected motion of the associated galaxy is pointing to the East.

\subsubsection{J1254-2901b (A\,3528~N)}\label{sec:j2901b}
The radio-optical overlay of this narrow-angle tailed radio galaxy is shown in the top left panel of Fig. \ref{fig:opt-radio_tailA3528}. The radio galaxy is 
located $\sim4^{\prime}$ East of the BCG of A\,3528~N, and it is very small, barely exceeding the size of the
optical galaxy. It is very weak, with a radio power considerably lower than typical cluster tailed radio galaxies (Table \ref{tab:cluster}). 
It is not visible in the VLA 8.4 GHz data presented here, most likely due to primary beam attenuation.
The 235 MHz image does not show further extension of the tail. The orientation of the tails at 610 MHz  suggests a 
direction of the motion towards East, similar to J\,1254-2901a.

\subsubsection{J1254-2904 (A\,3528~N)}\label{sec:j2904}
This is a very peculiar radio source, whose morphology is difficult to classify. The radio-optical overlay (top right panel of
Fig. \ref{fig:opt-radio_tailA3528}) clearly shows the core of the radio galaxy at 8.4 GHz, coincident with the 
optical counterpart (solid arrow in the figure), and suggesting an active nucleus. The rest of the radio 
emission extends southwards and could be interpreted as the lobes of a head-tail radio galaxy. At 610 MHz (white contours) the morphology
of the tail consists of three separate parts: a northern part, where the lobes are well visible, and well coincident with
the emission at 8.4 GHz, a bottleneck and a southern extension, almost perpendicular to the direction of the lobes (see top right panel in Fig. \ref{fig:opt-radio_tailA3528}). 
The overall extent 
of the radio emission is $\sim100^{\prime\prime}$, i.e. $\sim$90 kpc. This source is located in the region between
the centres of A\,3528~N and A\,3528~S, at a distance of $\sim4^{\prime}$ and $\sim10^{\prime}$ respectively.

An interesting possibility could be that the southern radio emission is actually part of another source. Indeed, Fig.~\ref{fig:opt-radio_tailA3528} (top right panel) clearly shows that 8.4 GHz emission is also associated with 
a $z=0.0704$ galaxy (RA$_{\rm J2000}=12^h54^m24.35^s$,  DEC$_{\rm J2000}=-29^{\circ}04^{\prime}52.6^{\prime\prime}$)
located just East of the southern extension (dashed arrow in the top right panel of
Fig. \ref{fig:opt-radio_tailA3528}). Some of the other nearby galaxies are located ad a similar redshift,
indicating the presence of a background group at $z\sim0.07$ which could produce the diffuse southern emission we see at 235 and 610 MHz.

\subsubsection{J1254-2916 (A\,3528~S)}\label{sec:j2916}
This tailed radio galaxy, shown in the bottom panel of Fig. \ref{fig:opt-radio_tailA3528}, is located 
$\sim3^{\prime}$ South of the BCG in A\,3528~S (e.g. J1254-2913, see section \ref{sec:j2913} and 
top right  panel of Fig. \ref{fig:opt-radio_bcgA3528}). At low frequencies the tail extends for 
$\sim80^{\prime\prime}$ (i.e. $\sim$ 80 kpc), and it is longer than at high frequencies.
The head of the radio galaxy points approximately to North-East, providing further support to the 
overall ``cluster weather'' towards E-NE seen both in A\,3528~N and in A\,3528~S.
The 8.4 GHz radio contours in the overlay show that the two jets bend abruptly just out of the optical 
counterpart, most likely at the transition between the interstellar and intergalactic medium.

\section{The radio galaxy population in the A\,3558 Complex}\label{sec:a3558}
The properties of the radio galaxy population in the A\,3558 complex were studied in detail at 1.4
GHz and 2.3 GHz with ATCA and VLA observations in \cite{venturi+97,venturi+00}, and \cite{giacintucci+04}. The most interesting  result of those earlier works is
the remarkable lack of radio galaxies in A\,3558, which affects the radio luminosity function of the 
population of elliptical galaxies in the whole cluster complex.
Here we focus our attention on the radio emission of the BCGs and of the tailed radio galaxies in 
the three Abell clusters, while we present their spectral analysis in Sect. \ref{sec:spx}.

The sample under study here is formed three BCGs, i.e. J\,1324-3140 (in A\,3556), J\,1327-3129b 
(in A\,3558) and J\,1333-3141 (in A\,3562) and two tailed radio galaxies, i.e. J\,1324-3138 (in A\,3556)  
and J\,1333-3141 (in A\,3562). 
The tailed radio galaxies have already been studied in detail in \cite{venturi+98,venturi+03}. Here we report our earlier results 
(see Sects. \ref{sec:j3138} and \ref{sec:j3141}) for completeness and include them in our overall discussion. The properties of the galaxies are reported in Table \ref{tab:cluster}. 

\begin{figure*}[h!]
\centering
{\includegraphics[width=0.45\textwidth]{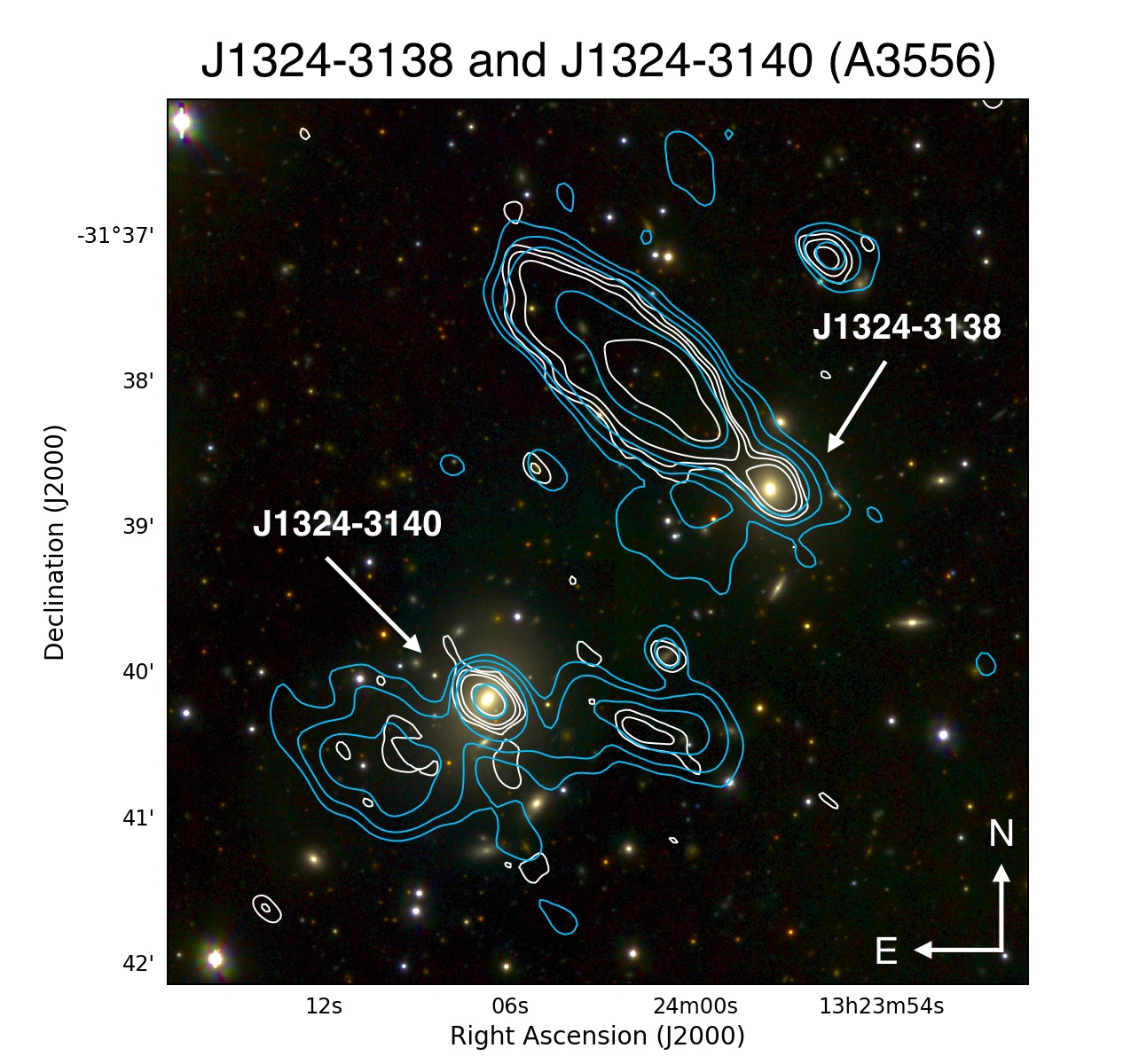}}
\hspace{2mm}
{\includegraphics[width=0.45\textwidth]{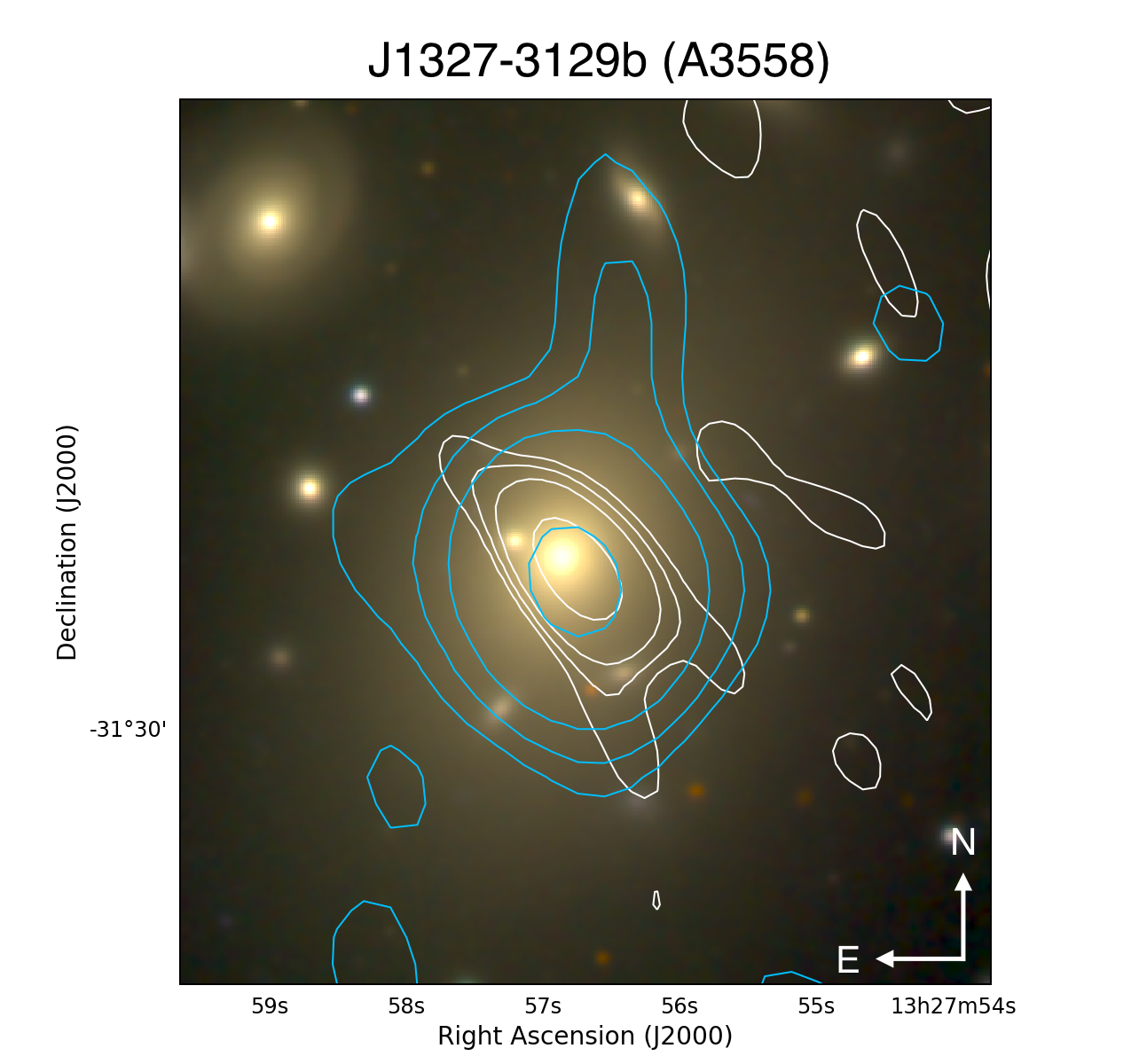}}
\hspace{2mm}
{\includegraphics[width=0.45\textwidth]{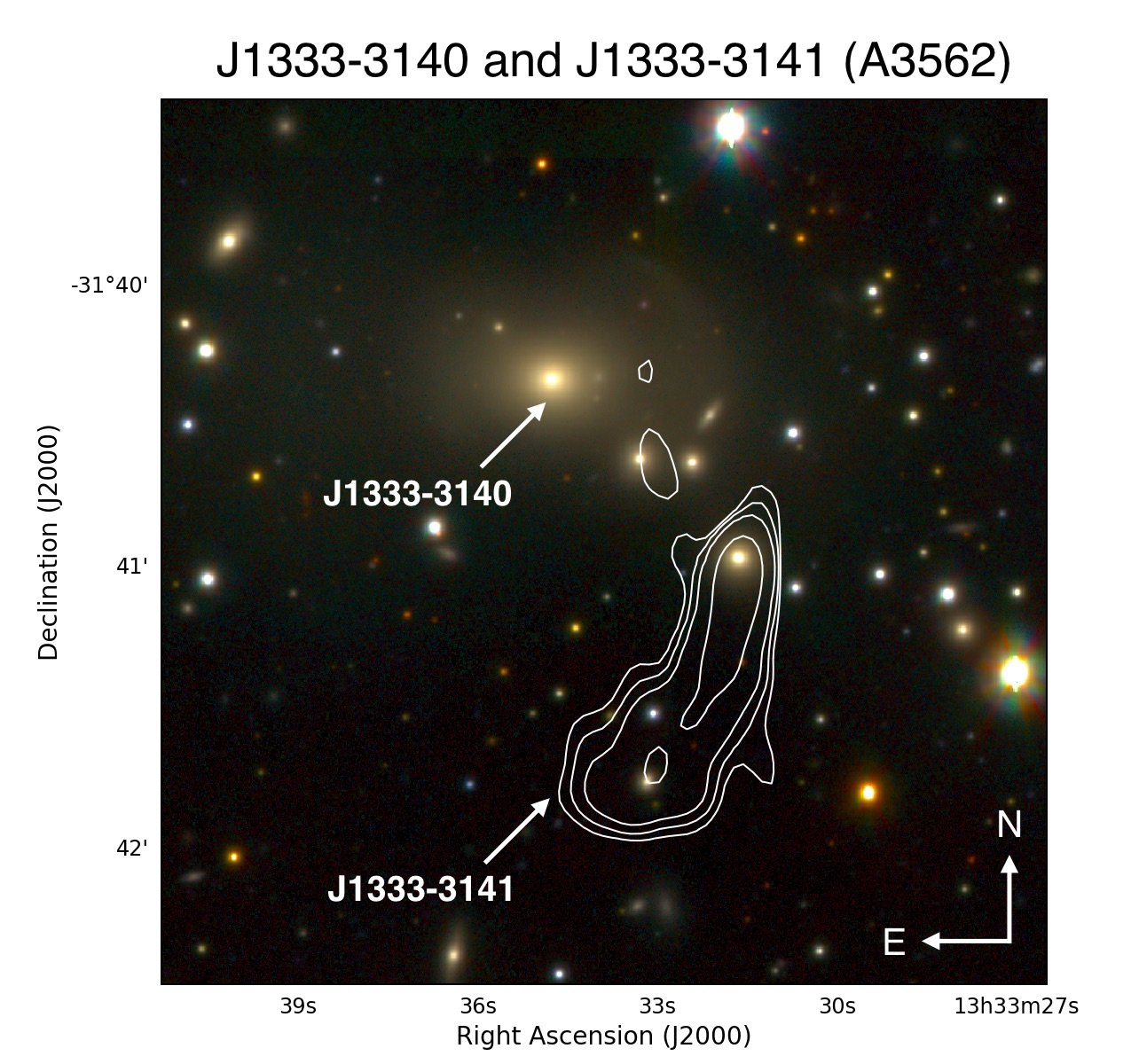}}
\caption{VST-ACESS {\it gri} composite images with GMRT radio contours at 325 (blue) and 610 (white) MHz. {\it Top left panel:} J\,1324-3138 and J\,1324-3140 (A\,3556). The 325 MHz contours are at $3\sigma\times(1, 2, 4, 8, 16)$ mJy beam$^{-1}$, the resolution is $16.3^{\prime\prime}\times11.89^{\prime\prime}$, p.a. $12.29^\circ$ and noise level $\sigma_{\rm 325~MHz}=0.20$ mJy beam$^{-1}$. The 610 MHz contours are at $3\sigma\times(2, 4, 8, 16)$ mJy beam$^{-1}$, the resolution is $12.86^{\prime\prime}\times7.2^{\prime\prime}$, p.a. $41.78^\circ$ and noise level $\sigma_{\rm 610~MHz}=0.08$ mJy beam$^{-1}$. {\it Top right panel:} J\,1327-3129b (A\,3558). The 325 MHz contours are at $3\sigma\times(2, 4, 8, 16)$ mJy beam$^{-1}$, the resolution is $16.3^{\prime\prime}\times11.89^{\prime\prime}$, p.a. $12.29^\circ$ and noise level $\sigma_{\rm 325~MHz}=0.25$ mJy beam$^{-1}$. The 610 MHz contours are at $3\sigma\times(2, 4, 8, 16)$ mJy beam$^{-1}$, the resolution is $10.98^{\prime\prime}\times5.59^{\prime\prime}$, p.a. $34.16^\circ$ and noise level $\sigma_{\rm 610~MHz}=0.14$ mJy beam$^{-1}$. {\it Bottom panel:} J\,1333-3140 and J\,1333-3141 (A\,3562). The 610 MHz contours are at $3\sigma\times(2, 4, 8, 16)$ mJy beam$^{-1}$, the resolution is $8^{\prime\prime}\times6^{\prime\prime}$, p.a. $0^\circ$ and noise level $\sigma_{\rm 610~MHz}=0.08$ mJy beam$^{-1}$.}
\label{fig:opt-radio_A3558}
\end{figure*}

\subsection{Radio properties of the BCGs}\label{sec:bcg_a3558}

\subsubsection{J1324-3140 (A\,3556)}\label{sec:j3140}
The radio emission associated with the BCG in A\,3556 is shown the the left panel of 
Fig. \ref{fig:opt-radio_A3558}, overlaid with an optical image. J\,1324-3140 is a double radio galaxy,
with a pair of symmetric lobes which become clearly visible only at 235 MHz and 325 MHz. Only 
hints of emission from the lobes are visible at 610 MHz, suggesting a steep synchrotron spectrum. 
The ATCA observations at 1.4 GHz presented in \cite{venturi+98} show low surface brightness
emission West of the nuclear component, in the region of the western lobe, which we were not able
to interpret at the time.
In the light of our GMRT images we can now associate this emission with the western lobe of the radio galaxy.

\subsubsection{J1327-3129b (A\,3558)}\label{sec:j3129b}
The radio emission associated with the BCG in A\,3558 is shown in the right panel of Fig. \ref{fig:opt-radio_A3558}. 
It is a compact and unresolved source at all frequencies of our observations.
The radio emission is fully confined into the faint optical galaxy halo. It is the weakest radio galaxy in our 
sample, as clear from Table \ref{tab:cluster}.

\subsubsection{J1333-3140 (A\,3562)}\label{sec:j33-3140}
The cluster A\,3562 was studied in detail in \cite{venturi+03} and \cite{giacintucci+05}, by means of GMRT and 
ATCA observations from 235 MHz to 8.4 GHz. The cluster hosts a radio halo, whose origin has been interpreted in
the light of the ongoing merging events in the region between A\,3558 and A\,3562. 
Our 325 MHz image (Fig. \ref{fig:mosaic_A3558}) is consistent with our earlier studies \citep{venturi+00} and confirm that the BCG in A\, 3562 is radio quiet (see also the optical-radio overlay in the bottom panel in Fig. \ref{fig:opt-radio_A3558}. The white arrow indicates the BCG in A3562.).

\subsection{Tailed radio galaxies}\label{sec:tail_a3558}
\subsubsection{J1324-3138 (A\,3556)}\label{sec:j3138}
J\,1324-3138 is shown in the left panel of Fig. 5. It lies very close to the centre of A\,3556, at a projected distance of
only $\sim 2.5^{\prime}$ from the BCG (e.g. J\,1324-3140). It was studied in detail  in \cite{venturi+98}, where they concluded that
the source is most likely a head-tail radio galaxy whose nuclear emission has switched off, based on the spectral analysis.
None of our observations is able to resolve the nuclear component and the inner jets (if present). 
Our new GMRT observations show a similar extent and morphology of those earlier studies,
ruling out further extended emission at low frequency.

\subsubsection{J1333-3141 (A\,3562)}\label{sec:j3141}
This head-tail radio galaxy was studied in detail from 235 MHz to 8.4 GHz with GMRT and ATCA 
observations \citep{venturi+03,giacintucci+05}. It is located  at a projected distance 
of $\sim 1^{\prime}$ South of the centre of A\,3562, the cluster that lies at the western 
end of the A\,3558 cluster complex (see Figure \ref{fig:mosaic_A3558}). The length of the tail is 
$\sim 1^{\prime}$ (i.e. $\sim60$ kpc at cluster's redshift), before it merges in the radio halo. 
It is not straight, but it smoothly bends from East to West, suggesting a possible orbital motion 
around the cluster BCG (see Fig. 6 in \citealt{venturi+03} and bottom panel in Fig. \ref{fig:opt-radio_A3558}). No further imaging or analysis is presented in this paper, but we review 
the literature information here to help the discussion which will be presented in Section \ref{sec:disc}.

%
\begin{table*}[h!]
\caption[]{Flux measurements for the sample of the BCGs. The information on the observations (e.g. beam and literature references) are settled in Tables \ref{tab:array_res} and \ref{tab:archive}.}
\centering
\begin{tabular}{cccccccc}
\hline
\hline\noalign{\smallskip}
Radio & Cluster &Telescope &  $\nu$ & $S_{\rm tot}$ & $S_{\rm central}$ & $S_{\rm diff}$  & Ref.\\
galaxy & name &  & (MHz) & (mJy) & (mJy) & (mJy) \\
 \hline\noalign{\smallskip}
\multirow{5}{*}{J\,1254-2900} &  \multirow{5}{*}{A\,3528~N}	& GMRT & 235 & $996.1\pm79.7$  & & & this work \\
							& & GMRT & 610 & $428.1\pm21.4$  & & & this work \\
							& & ATCA & 1380 & $230.9\pm6.9$ & & & \citealt{venturi+01}\\
							& & ATCA & 2380 & $142.4\pm4.3$ & & & \citealt{venturi+01} \\
							& & VLA & 8400 & $50.5\pm1.5$  & & & this work \\ 
\hline\noalign{\smallskip}
\multirow{8}{*}{J\,1254-2913} & \multirow{8}{*}{A\,3528~S}	& GMRT & 150 & $14491.0\pm2898.2$ &  & & TGSS image  \\
							& & GMRT& 235 & $7291.4\pm583.3$ & $4714.5\pm377.2$ & $2576.9\pm694.6$  & this work\\
							& & GMRT& 610 & $2159.1\pm108.0$ & $1956.7\pm97.8$ & $202.4\pm145.8$  & this work \\
							& & MOST & 843 & $1970.0\pm59.1$ &  &  & SUMSS image \\
							& & ATCA & 1380 &  & $936.7\pm28.1$ & & \citealt{venturi+01} \\
							& & VLA & 1400 &  $1069.1\pm32.1$ &  & $132.4\pm42.6^\star$ & NVSS image \\
							& &  ATCA & 2380 &  & $538.4\pm16.2$ & &  \citealt{venturi+01} \\
							& & VLA & 8400 &  & $125.3\pm3.8$ & & this work \\
\hline\noalign{\smallskip}
\multirow{7}{*}{J\,1257-3021} & \multirow{7}{*}{A\,3532}	& GMRT & 150 & $8655.9\pm1410.1$ & $6802.4\pm1360.5$ & $1853.5\pm370.7$ & TGSS image \\
							& & GMRT & 235 & $4591.9\pm336.0$ & $4251.3\pm334.9$ & $340.5\pm27.2$ & this work \\
							& & GMRT & 610 & $1522.0	\pm74.2$ & $1484.0\pm74.2$ & $38.0\pm1.9$ & this work \\
							& & MOST & 840  & $1773.0\pm53.2$ & & & SUMSS image \\
							& & ATCA & 1380 & & $1056.5\pm31.7$ &  & \citealt{venturi+01}\\
							& & VLA & 1400  & $1061.0\pm31.8$ & $4.5\pm44.4^{\star\diamond}$ & & NVSS image\\
							& & ATCA & 2380 & & $651.7\pm19.6$ & & \citealt{venturi+01} \\
\hline\noalign{\smallskip}
\multirow{7}{*}{J\,1324-3140} & \multirow{7}{*}{A\,3556}	& GMRT & 235 & $77.1\pm6.2$ & $18.6\pm1.5$ & $58.5\pm6.4$ & this work\\
							& & GMRT & 325 & $84.6\pm4.2$ & $16.8\pm0.8$ & $67.8\pm4.3$ & this work \\
							& & GMRT & 610 & $18.7\pm0.9$ & $10.9\pm0.5$ & $7.8\pm1.0$ & this work \\
							& & MOST & 843 & $16.7\pm0.5$ & & & SUMSS image\\
							& & ATCA & 1380 &  & $7.8\pm0.2$ &  & \citealt{venturi+97}\\
							& & ATCA & 2380 &  & $7.4\pm0.2$ &  & \citealt{venturi+97} \\
							& & ATCA & 4790 &  & $3.0\pm0.1$ &  & \citealt{venturi+08}\\
\hline\noalign{\smallskip}
\multirow{5}{*}{J\,1327-3129b} & \multirow{5}{*}{A\,3558}	& GMRT & 235 & $19.1\pm1.5$ & & & this work\\
							& & GMRT & 325 & $17.9\pm1.4$ & & & this work\\
						 	& & GMRT & 610 & $14.6\pm0.7$ & &  & this work\\
							& & ATCA & 1380 & $6.2\pm0.2$ & & &\citealt{venturi+97}\\
							& & ATCA & 2380 & $1.5\pm0.1$ & & & \citealt{venturi+97}\\
\hline\noalign{\smallskip}
\end{tabular}
\tablefoot{The flux density of the diffuse emission is the difference between the total flux density and the central
emission, i.e. the emission contained within the optical counterpart, which is encompassed in the region highlighted
in red in Figs. \ref{fig:spectra_bcg} and \ref{fig:spectra_tail}.  The resolution of TGSS does not allow
to separate J\,1254-2900 and J\,1254-2901a (A\,3528~N).
J\,1254-2900 (A\,3528~N) and J\,1327-3129b (A\,3558) do not show extended emission, and we provide only the total flux density. 
For J\,1257-3021 (A\,3532) we measured the flux density of the diffuse and central emission separately, and the total one as a sum of the two.
$^\star$ The diffuse emission was calculated by difference between the VLA (total region) and the ATCA (central region) 
observation. $^\diamond$ Upper limit.}
\label{tab:flux_bcg}
\end{table*}

\begin{table*}[h!]
\caption[]{Flux measurements for the sample of the tailed galaxies. The information on the observations (e.g. beam and literature references) are settled in Tables \ref{tab:array_res} and \ref{tab:archive}.}
\centering
\begin{tabular}{cccccccc}
\hline
\hline\noalign{\smallskip}
Radio & Cluster & Telescope &  $\nu$ & $S_{\rm tot}$ & $S_{\rm c}$ & $S_{\rm diff}$ & Ref. \\
galaxy & name & & (MHz) & (mJy) & (mJy) & (mJy) \\
 \hline\noalign{\smallskip}
\multirow{5}{*}{J\,1254-2901a} & \multirow{5}{*}{A\,3528~N} & GMRT & 235 & $349.4\pm28.0$ & $239.9\pm19.2$ & $109.5\pm34.0$ & this work\\
							& & GMRT & 610 & $186.6\pm9.3$ & $140.0\pm7.0$ & $46.6\pm11.6$ & this work\\
							& & ATCA & 1380 & & $110.5\pm3.3$ & & \citealt{venturi+01}\\
							& & ATCA & 2380 & & $57.9\pm1.7$ & & \citealt{venturi+01}\\
							& & VLA & 8400 & & $22.1\pm0.7$ & & this work\\ 
\hline\noalign{\smallskip}
\multirow{2}{*}{J\,1254-2901b} & \multirow{2}{*}{A\,3528~N} 	& GMRT & 235 & $22.7\pm1.8$ & & & this work \\
							& & GMRT & 610 & $13.1\pm0.7$ & & & this work\\
\hline\noalign{\smallskip}
\multirow{8}{*}{J\,1254-2904} & \multirow{8}{*}{A\,3528~N}	& GMRT & 150 & $1463.7\pm292.7$ & $607.4\pm121.5$ & $856.3\pm316.9$ & TGSS image \\
							& & GMRT & 235 & $926.6\pm74.1$ & $416.1\pm33.3$ & $510.6\pm81.2$  & this work\\
							& & GMRT & 610 & $342.8\pm17.1$ & $221.4\pm11.1$ & $121.4\pm20.4$  & this work\\
							& & MOST & 843 & $519.0\pm15.6$ &  & & SUMSS image \\
							& & ATCA & 1380 & $295.9\pm8.9$ & & & \citealt{venturi+01} \\
							& & VLA & 1400 & $311.7\pm9.4$ & & & NVSS image  \\
							& & ATCA & 2380 & $142.8\pm4.3$ & 126.2$\pm$3.8 & 16.6$\pm5.7$  & \citealt{venturi+01} \\
							& & VLA & 8400 & & 30.0$\pm$1.0 & & this work \\
\hline\noalign{\smallskip}
\multirow{7}{*}{J\,1254-2916} & \multirow{7}{*}{A\,3528~S}	& GMRT& 235 & $259.3\pm20.7$ & $132.8\pm10.6$ & $126.5\pm23.3$  & this work \\
							& & GMRT& 610 & $149.1\pm7.5$ & $76.5\pm3.8$ & $72.6\pm8.4$ & this work \\
							& & MOST & 843 & $124.0\pm3.7$ &  & & SUMSS image  \\
							& & ATCA & 1380 & & $62.1\pm1.9$ & & \citealt{venturi+01} \\
							& & VLA & 1400 & $85.8\pm2.6$ &  & $23.7\pm3.2^\star$ & NVSS image\\
							& & ATCA & 2380 & & $49.6\pm1.5$ &  & \citealt{venturi+01}  \\
							& & VLA & 8400 & & $15.3\pm0.5$ &  & this work \\
\hline\noalign{\smallskip}
\multirow{11}{*}{J\,1324-3138} & \multirow{11}{*}{A\,3556} & GMRT & 235 & $278.3\pm22.3$ & $9.1\pm0.7$ & $269.2\pm22.3$ & this work \\
						   & & VLA & 327 & $230\pm11.5$ &	& & \citealt{venturi+98}\\
						   & & GMRT & 325 & $254\pm12.7$ & $8.1\pm0.4$ & $245.9\pm12.7$  & this work\\
						   & & GMRT & 610 & $122.7\pm6.1$ & $6.2\pm0.2$ & $119.5\pm6.1$ & this work \\
						   & & MOST & 843 & $80.2\pm2.4$ & & &\citealt{venturi+98}\\
						   & & ATCA & 1380 & $41\pm1.2$ & $2.1\pm0.1$ & $38.9\pm1.2$ & \citealt{venturi+97}\\
						   & & ATCA & 1400 & $43.1\pm1.3$ & $2.7\pm0.1$ & $40.4\pm1.3$ & ATCA archive image\\
						   & & ATCA & 2380 & $21.0\pm0.6$ & $1.7\pm0.1$ & $19.3\pm0.6$ & \citealt{venturi+97}\\
						   & & ATCA & 4790 & $7.3\pm0.3$ & $0.5\pm0.03$ & $6.8\pm0.3$ & \citealt{venturi+98}\\
						   & & ATCA & 5000 & $6.8\pm0.2$ & $0.4\pm0.01$ & $6.5\pm0.2$  & ATCA archive image\\
						   & & ATCA & 8640 & $2.2\pm0.1$ & $0.1\pm0.01$ & $2.1\pm0.1$ & \citealt{venturi+98}\\
\hline\noalign{\smallskip}
\end{tabular}
\tablefoot{The diffuse emission is the difference between the total flux density and the emission from the central components.
$^\star$The diffuse emission was calculated by difference between the VLA (total region) and the ATCA (central region) observation.}
\label{tab:flux_tail}
\end{table*}

\begin{figure*}[h!]
\centering
{\includegraphics[width=0.40\textwidth]{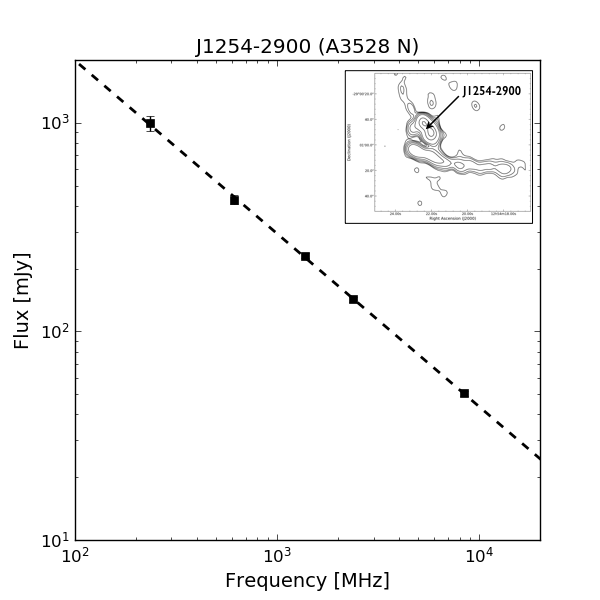}}
\hspace{0mm}
{\includegraphics[width=0.40\textwidth]{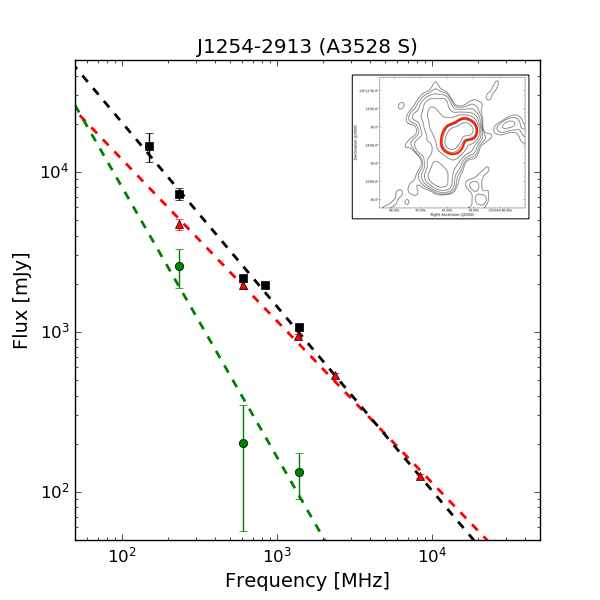}}\\
\smallskip
{\includegraphics[width=0.40\textwidth]{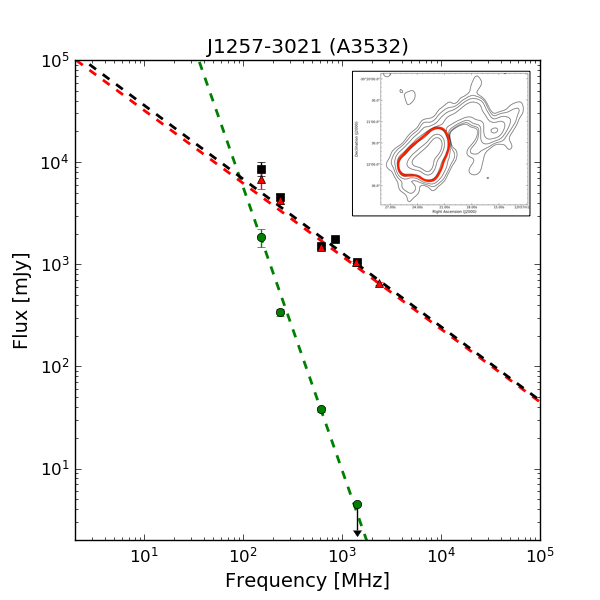}}
\hspace{0mm}
{\includegraphics[width=0.40\textwidth]{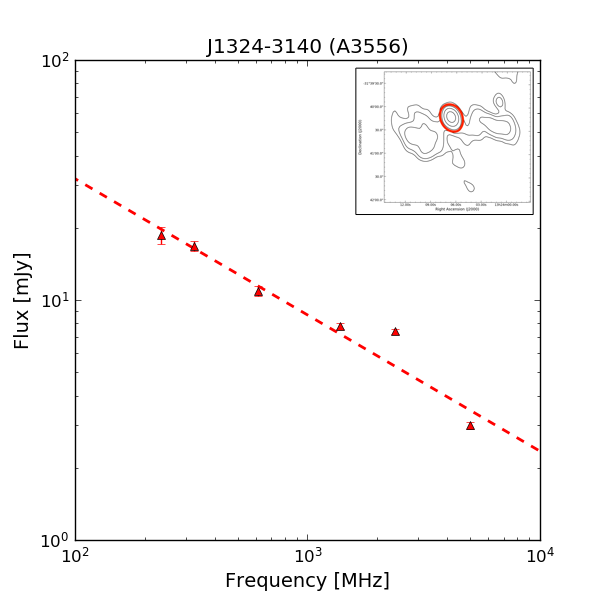}}\\
\smallskip
{\includegraphics[width=0.40\textwidth]{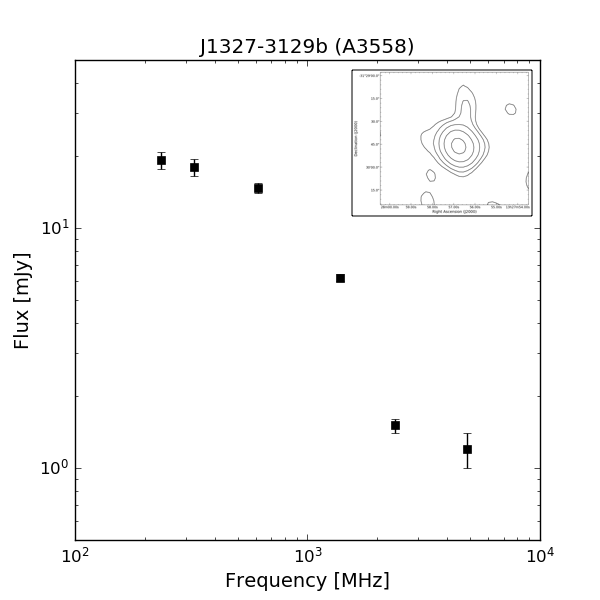}}
\caption{Integrated radio spectra for the total (black), central (red) and diffuse (green) emission for the BCGs sample. The contours of the radiogalaxy are displayed in the top right corner of each panel. The red line shows the region that has been taken as ``central'' emission. The value at 1.4 GHz in the J\,1257-3021 (A\,3532) spectrum is un upper limit.}
\label{fig:spectra_bcg}
\end{figure*}

\begin{figure*}[h!]
\centering
{\includegraphics[width=0.40\textwidth]{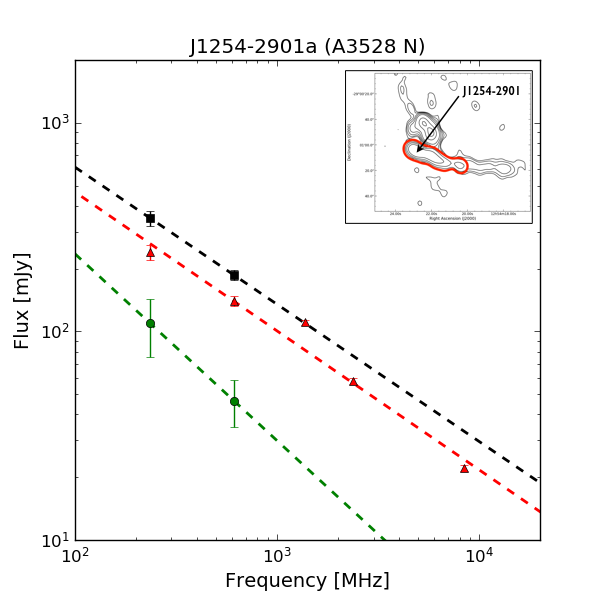}}
\hspace{0mm}
{\includegraphics[width=0.40\textwidth]{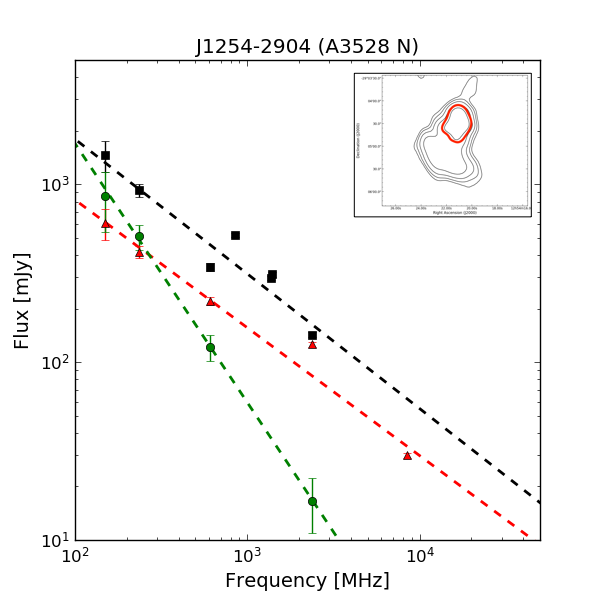}}\\
\smallskip
{\includegraphics[width=0.40\textwidth]{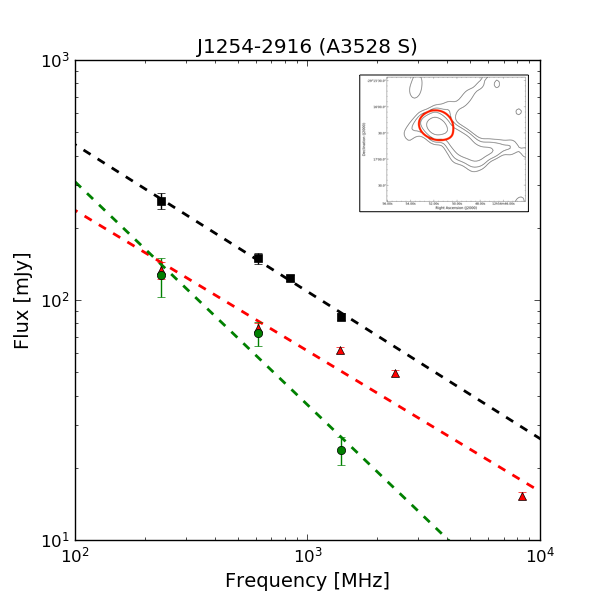}}
\hspace{0mm}
{\includegraphics[width=0.40\textwidth]{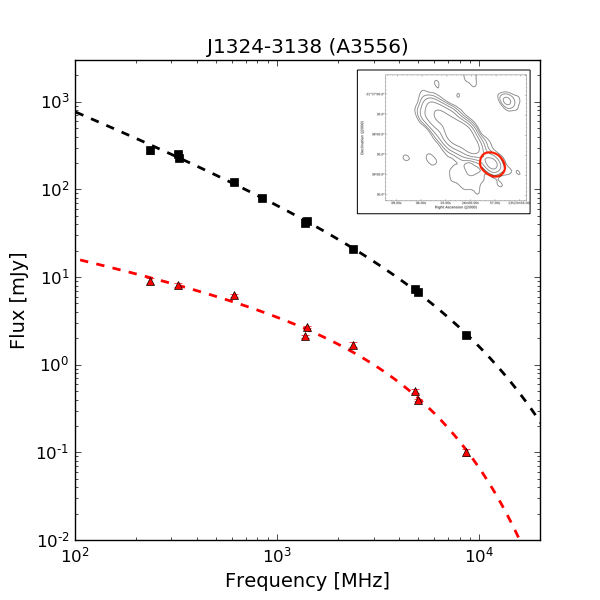}}\\
\caption{Integrated radio spectra for the total (black), central (red) and diffuse (green) emission for the tails sample. The contours of the radiogalaxy are displayed in the top right corner of each spectrum. The red line shows the region that has been taken as ``central'' emission}
\label{fig:spectra_tail}
\end{figure*}

\section{Spectral analysis}\label{sec:spx}

In order to study the nature of the radio emission of the galaxies in our sample, and possibly identify
different evolutionary stages in their lifecycle, we performed a spectral study of each radio galaxy
in the frequency range 150 MHz-8.46 GHz, by means of the information at six to nine different 
frequencies (see Tables \ref{tab:flux_bcg} and \ref{tab:flux_tail}).
Our procedure and results are reported in the next subsections. The spectra and best fit results 
from radiation loss models for the BCGs and for the tailed radio galaxies are reported in Fig. 
\ref{fig:spectra_bcg} and \ref{fig:spectra_tail}.

\begin{table*}[h!]
\caption[]{Best-fit model of the radiative losses of the central region of the radio galaxies and their physical parameters.}
\centering
\renewcommand{\arraystretch}{1.5}{%
\begin{tabular}{cccccccccc}
\hline\hline
Radio & Cluster &comp. & rad. loss & $\alpha_{\rm inj}$ & $\chi^2_{\rm red}$ & $\nu_{\rm br}$ &    $B_{\rm eq}$ & $P_{\rm int}$        & $t_{\rm rad}$ \\
galaxy & name & & model & & & (GHz) & ($\mu{\rm G}$) & (dyne cm$^{-2}$) & (Myr) \\
\hline
J\,1254-2900 & A\,3528~N & tot.      & PL & $0.83^{+0.03}_{-0.11}$ & 0.11 & $\geq 8.4$ & 10.6 & $6.4\times10^{-12}$ & 14\\
\hline
J\,1254-2901a 	& A\,3528~N & centr. & PL & $0.67^{+0.03}_{-0.10}$ & 0.7 & $\geq 8.4$ & 7.2 & $3.1\times10^{-12}$ & 22\\
\hline
\multirow{3}{*}{J\,1254-2904} & \multirow{3}{*} {A\,3528~N} & centr. &PL & $0.72^{+0.03}_{-0.08}$ & 1.1 & $\geq 8.4$ & 4.7 & $1.2\times10^{-12}$ &  34 \\
		      			    & & diff.    & PL & $1.45^{+0.05}_{-0.21}$ & 0.06 & $\geq 8.4$ & 3.5 & $7.2\times10^{-13}$ & 41 \\
					    & & tot.    & PL  & $0.76^{+0.03}_{-0.13}$ & 1.5 & $\geq 8.4$ & 4.1 & $9.6\times10^{-13}$ & 38  \\
\hline
\multirow{3}{*}{J\,1254-2913} & \multirow{3}{*} {A\,3528~S} & centr. & PL & $1.01^{+0.03}_{-0.10}$ & 0.3 & $\geq 8.4$ & 8.1 & $3.8\times10^{-12}$ & 19 \\
		      				& & diff.    & PL & $1.77^{+0.65}_{-0.56}$ & 3.1 & $\geq 8.4$ & 10.0 & $5.7\times10{-12}$ & 15 \\
		      				& & tot.    & PL & $1.15^{+0.03}_{-0.19}$ & 0.6 & $\geq 8.4$ & 10.2 &   $6.1\times10^{-12}$ &  15 \\
\hline
\multirow{3}{*}{J\,1254-2916} & \multirow{3}{*}{A\,3528~S} & cen. & PL & $0.59^{+0.06}_{-0.10}$ & 1.1 & $\geq 8.4$ & 5.2 & $1.6\times10^{-12}$  & 31\\
						& & diff.    & PL & $0.93^{+0.21}_{-0.22}$ & 1.7 & $\geq 8.4$ & 3.9 & $8.7\times10^{-13}$ & 40 \\
						& & tot      & PL & $0.61^{+0.03}_{-0.24}$ & 0.03 & $\geq 8.4$ & 4.3 & $1.1\times10^{-12}$  & 36\\
\hline
\multirow{3}{*}{J\,1257-3021} & \multirow{3}{*}{A\,3532}  & centr. & PL & $0.71^{+0.02}_{-0.06}$ & 7.4 & $\geq 8.4$ & 6.8 & $2.7\times10^{-12}$ & 24 \\
		      			  	& & diff.     & PL & $2.63^{+0.43}_{-0.37}$ & 1.6 & $\geq 8.4$ & 4.9 & $1.4\times10^{-12}$ &  33 \\
						& & tot.    & PL & $0.72^{+0.01}_{-0.09} $ & 13 &  $\geq 8.4$  & 6.7 & $2.6\times10^{-12}$ &  24 \\
\hline
\multirow{2}{*}{J\,1324-3138}  & \multirow{2}{*}{ A\,3556} & centr. & JP & 0.5* & 6.7 & $3.9^{+0.1}_{-0.2}$ & 3.4 & $6.7\times10^{-13}$ & 62\\
		     				 & & diff.    & JP & $0.95^{+0.03}_{-0.07}$ & 2.1 & $11.2^{+1.3}_{-2.6}$ & 2.9 & $4.9\times10^{-13}$ & 39\\
\hline
J\,1324-3140  &  A\,3556 & centr. & PL & $0.57^{+0.04}_{-0.11}$ & 0.7 & $\geq 8.4$ & 5.2 & $1.5\times10^{-12}$ & 32 \\
\hline
\end{tabular}}
\tablefoot{* fixed $\alpha$ value. For all the PL models we obtain an upper limit estimation of the radiogalaxy lifetime.}
\label{tab:nu_br}
\end{table*}

\subsection{Integrated spectra}\label{sec:int_spectra}

The total integrated spectrum for all galaxies in the sample was derived using the flux density 
measurements reported in 
Tables \ref{tab:flux_bcg} and \ref{tab:flux_tail}. The tables report three sets of measurements. 
In particular: 

\begin{itemize}
\item{}The total flux density measurements at each frequency were obtained by means of the AIPS 
task \texttt{TVSTAT} 
integrating within  the $3\sigma$ contour level. We took the 235 MHz emitting volume as the 
reference, and we measured 
the flux density at each frequency over the same volume, to account for the different sensitivities of 
our images.
The only exception is J\,1324-3140 (BCG in A\,3556) for which we used the best available image at 325 MHz;
\medskip
\item{} we refer to the flux density of the ``central'' region, meaning the emission which is clearly associated
with the optical counterpart. We are aware that this definition is not very rigorous. In a couple of cases, as in J\,1324-3138 
and J\,1324-3140 (radio tail and BCG in A\,3556, respectively),  it is coincident with the radio core, while in other cases it refers to the full radio 
emission except for the extended component, which is clearly separated, as in the case of J\,1254-2901a and J\,1254-2904 (radio tails in A\,3528~N),
J\,1254-2913 (BCG in A\,3528~S) and J\,1257-3021 (BCG in A\,3532). The strong edge brightening of the latter two sources provides a natural definition
of this part of the radio emission.
We point out that due to the average resolution of our images, in most cases the compact core is clearly imaged 
only at 8.4 GHz, as is the case of J\,1254-2904 (radio tail in A\,3528~N), J\,1254-2913 and J\,1254-2916 (BCG and radio tail in A\,3528~S, respectively), while it blends with the rest
of the ``central'' region at any other frequency. For this reason a spectral study of the core is not feasible;
\medskip
\item{} finally, the flux density of the diffuse emission is obtained by subtraction of the central flux density from the total one. 
The only exceptions are J\,1257-3021 ( BCG in A\,3532) and J\,1324-3140  (BCG in A\,3556), where the values for the central and the diffuse component 
were obtained individually, as they can be easily separated in our images. 

\end{itemize}

For better clarity, the insets in Fig. \ref{fig:spectra_bcg} and  \ref{fig:spectra_tail} show the region corresponding to total emission (black contours) and to the central region (single red contour).
The flux density error is derived as the sum of the contribution of the thermal noise of the image and of the residual calibration 
error. The latter is dominating in our cases \citep{chandra+04}.

\subsubsection{Spectral properties of the BCGs}\label{sec:int_spectra_bcg}

Here below we provide further details for each source.

\begin{itemize}

\item {\bf J\,1254-2900 (A\,3528~N) - } No diffuse emission is detected for this radio galaxy (see Sect. \ref{sec:j2900}). The
spectrum is well fitted by a single power law with $\alpha^{\rm 8.4~GHz}_{\rm 235~MHz}=0.83^{+0.03}_{-0.11}$, typical of an active 
source emitting via synchrotron radiation. 
\medskip

\item {} {\bf J\,1254-2913 (A\,3528~S) - }  The spectrum of the central region is well fitted by a single power law, with 
$\alpha^{\rm 8.4~GHz}_{\rm 235~MHz}=1.01^{+0.03}_{-0.11}$.
The spectral index of the total emission up to 610 MHz was obtained both with and without the 150 MHz flux density 
measurement. In both cases we obtained $\alpha\sim 1.2$. Since the total emission is dominated by the central
region, the similarity of the two spectra is not surprising.
The spectral index of the diffuse emission is steeper, with $\alpha^{\rm 1.4~GHz}_{\rm 235~MHz}=1.77^{+0.65}_{-0.56}$.
We note that our flux density measurement at 610 MHz is well below this power-law.
Even considering this value a lower limit (in the light of the hints of extended emission at the sensitivity level of our observations), the difference between the spectrum of the central emission,
the steep value of the diffuse emission and the sharp morphological transition and surface brightness distribution between the central and diffuse part of the radio emission cannot be
ignored.

\medskip

\item {} {\bf J\,1257-3021 (A\,3532) -} The spectrum of the central region of the radio galaxy is consistent with an active structure ($\alpha^{\rm 2.38~GHz}_{\rm 235~MHz}=0.71^{+0.02}_{-0.06}$). The spectral index of the diffuse emission in the North-West direction, observed at 150, 235 and 610 MHz is $\alpha^{\rm 610~MHz}_{\rm 150~MHz}=2.63^{+0.43}_{-0.37}$. 
As in the case of J\,1254-2913 (in A\,3528~S) the sharp morphological transition and surface brightness drop between the
central region and the diffuse emission is noticeable.

\medskip

\item {} {\bf J\,1324-3140 (A\,3556) -} Beyond the compact component associated with the nucleus of the BCG, the low frequency 
observations show the presence of a pair of symmetric lobes, which are well imaged only at 325 MHz, most likely due to the
fact that the u-v coverage of our snapshot observations at 610 MHz and 235 MHz is inadequate to image such low
brightness features. This is suggested by the scatter in the flux density of the lobes between 235 MHz and 610 MHz, which does not allow any estimate of the spectral index, and the flux density
values  at these two frequencies should be considered lower limits.
The spectral behaviour of the central region is very different. Albeit some scattering, its spectrum is very flat, i.e.,
$\alpha^{\rm 5~GHz}_{\rm 235~MHz}=0.57^{+0.04}_{-0.11}$.
It is possible that the nuclear region of J\,1324-3140 is variable:
our measurements span a time interval of almost 30 years, and variability in the source would 
indeed reproduce the observed scatter. 
We point out that calibration problems at any of the frequencies are ruled out by the consistency of
the flux density measurements of the other sources in the field, and in particular J\,1324-3138 (in A\,3556, see Sect. \ref{sec:int_spectra_tail}).

\medskip

\item {} {\bf J\,1327-3129b (A\,3558) -}  This source is all confined within the nucleus of the BCG, and its spectrum  
shows a peculiar trend. It is almost flat below 610 MHz, then it steepens dramatically above this frequency.
The spectral index has been calculated into three frequency intervals, i.e., 
$\alpha^{\rm 610~MHz}_{\rm 235~MHz}=0.28\pm0.07$, 
$\alpha^{\rm 1.4~GHz}_{\rm 610~MHz}=1.03\pm0.09$ and 
$\alpha^{\rm 2.4~GHz}_{\rm 1.4~GHz}=2.63\pm0.50$. 
We also re-analyzed archival VLA data at 4.85 GHz (CnB configuration, beam size of $\sim11^{\prime\prime}$ resolution).
At this frequency we measured a radio flux density of $1.2\pm0.2$ mJy, which led to a spectral index of $\alpha^{\rm 4.9~GHz}_{\rm 1.4~GHz}=1.31\pm0.14$.
Considering that the source is undetected in the TGSS, the overall spectrum is suggestive of a 
concave shape, typical of compact steep spectrum sources \citep[CSS, see ][]{fanti+95} or 
megahertz peaked sources \citep[MPS,][]{coppejans+17}, i.e. active galactic nuclei in the early
stages of their radio activity. The steep spectrum in the optically thin region suggests the presence of a population of old electrons, although a 
steeper spectrum in the optically thin region has been recently found for the ``young'' CSS J1613+4223 \citep[$\alpha\sim1.6$]{dallacasa+orienti16}.

\end{itemize}

\subsubsection{Spectral properties of the tailed radio galaxies}\label{sec:int_spectra_tail}
Following the same procedure as for the BCGs, we performed the spectral analysis for the tailed 
radio sources, by means of the integrated fluxes in Table \ref{tab:flux_tail}.

\begin{itemize}
\item {} {\bf J\,1254-2901a (A\,3528~N) -} The central region, which includes the core and inner part 
of the tail, has a spectrum with $\alpha^{\rm 8.4~GHz}_{\rm 235~MHz}=0.67^{+0.03}_{-0.10}$. 
At low frequencies the tail extends to $\sim$ 80 kpc (see Sect. 4.2.1) and  its spectrum steepens 
from $\alpha^{\rm 610~MHz}_{\rm 235~MHz}=0.56\pm0.10$ to 
$\alpha^{\rm 610~MHz}_{\rm 235~MHz}=0.89\pm0.42$, consistent  with what is usually seen in 
tailed radio galaxies \citep[e.g.,][]{pizzo+09,stroe+13}. 

\medskip

\item {} {\bf J\,1254-2901b (A\,3528~N) -} The spectrum of this tail, in the very limited range available here, is 
$\alpha_{\rm 235~MHz}^{\rm 610~MHz}=0.58\pm0.10$, again typical of an active source.

\medskip

\item {} {\bf J\,1254-2904 (A\,3528~N) -} The active nucleus associated with the optical counterpart is clearly
separated from the rest of the emission only at 8.4 GHz, hence no estimate of its spectral index can be derived 
from our images. The central region is detected all the way up to 8.4 GHz, with a spectral index 
$\alpha^{\rm 8.4~GHz}_{\rm 150~MHz}=0.72^{+0.03}_{-0.08}$.
Conversely the spectrum of the southern extension is considerably steeper, i,e. 
$\alpha^{\rm 2.38~GHz}_{\rm 150~MHz}=1.45^{+0.05}_{-0.21}$. The reason for the abrupt change
in the morphological and spectral properties of this radio galaxies South of the bottleneck is 
unclear. It is becoming evident that the interplay between the radio plasma in galaxy clusters and 
the presence of perturbations in the intracluster medium (such as for instance shocks) may be 
common \cite[i.e.,][]{shimwell+16,degasperin+17}. 
Considering that J\,1254-2904 is located in the region between A\,3528~N and A\,3528~S, the
possibility that the southern extension is old revived radio plasma is intriguing. 
An alternative possibility is that this region is actually aged radio plasma associated with the
$z=0.0704$ galaxy located
just outside the radio contours (see Sect. \ref{sec:j2904} and Fig. \ref{fig:opt-radio_tailA3528}).

\medskip

\item {} {\bf J\,1254-2916 (A\,3528~S) -} As for J\,1254-2901a  (in A\,3528~N), the length of the tail increases with
decreasing frequency (see Fig. \ref{fig:opt-radio_tailA3528}). At low frequencies (235 and 610
MHz)  the spectral index is $\alpha^{\rm 610~MHz}_{\rm 235~MHz}=0.98^{+0.21}_{-0.22}$. The central
region is well described by a power law,  
$\alpha^{\rm8.4~GHz}_{\rm235~MHz}=0.59^{+0.03}_{-0.06}$.

\end{itemize}

For the analysis of J\,1324-3138 (in A\,3556) and J\,1333-3141 (in A\,3562) we refer to literature works \citep[respectively]{venturi+98,venturi+03}. 
For J\,1324-3138 (in A\,3556) the flux density values provided by all the 
new GMRT observations presented here are in excellent  agreement with the spectrum reported 
in \citealt{venturi+98} and \citealt{venturi+03}, respectively (see Table 6, Sections \ref{sec:j3138} 
and \ref{sec:j3141}, Fig. \ref{fig:opt-radio_A3558}, \ref{fig:spectra_bcg} and \ref{fig:spectra_tail}).

\subsection{Spectral fits}\label{sec:fits}

We fitted the spectra of the central regions of the galaxies in our sample using the Synage++ package 
\citep{murgia+11}, to obtain an estimate of the radiative ages of the sources (subsection \ref{sec:rad_age}). 
We found that different radiative loss models were best suited in the various cases  
(see Table \ref{tab:nu_br}).  In particular:

\begin{itemize}

\item {} The spectra of J\,1254-2900, J\,1254-2901a and J\,1254-2904 (i.e. the BCG and the two radio tails in A\,3528~N), J\,1254-2913 and J\,1254-2916 (i.e. the BCG and the tail in A\,3528~S), J\,1257-3021 (i.e. the BCG in A\,3532) and J\,1324-3140 (i.e. the BCG in A\,3556) 
are best fitted by a power law; 

\smallskip

\item {} J\,1324-3138 (i.e. the radio tail in A\,3556) is best fitted by a Jaffe \& Perola model \citep[JP, ][]{jaffe+74}, which 
assumes continuous isotropisation of the pitch angle of the radiating electrons  by electron scattering, whose time scale is 
much shorter than their radiative lifetime; 

\smallskip

\item {} None of the above models provides an acceptable fit to the spectrum of J\,1327-3129b (i.e. the BCG in A\,3558), 
which we interpret as another piece of evidence in favour of the possibility that this is a compact steep spectrum source.

\end{itemize}

The best-fit values for the break frequency $\nu_{\rm br}$ are reported in Table \ref{tab:nu_br}. Note that for the
sources fitted with a power law, the break frequency is a lower limit, and the radiative age is an upper limit.

\subsubsection{Radiative ages and physical parameters}\label{sec:rad_age}
We used the break frequencies from the spectral fits (see Table \ref{tab:nu_br}) to estimate the radiative age of all the radio 
galaxies in the sample, under the assumption of equipartition \citep{govoni+04}, by means of the relation:

$$t_{\rm rad}=1590 \, [(1+z)\nu_{\rm br}]^{-1/2} \, \frac{B_{\rm eq}^{1/2}}{B_{\rm eq}^2+B^2_{\rm CMB}} \quad {\rm (Myr)} $$
where $B_{\rm CMB}=3.25(1+z)^2$. The magnetic field is expressed in $\mu$G, and the frequency in GHz. To estimate
$B_{\rm eq}$, i.e. the magnetic field at the equipartition, we assumed elliptical geometry (i.e.  $V=ab\pi/6$, with $a$ and $b$ the major and minor semi-axes of the ellipse, respectively),
 a filling factor of unity, and equal energy contribution in relativistic ions 
and electrons ($k=1$) to the total particle energy:

\begin{equation}
B_{\rm eq}=\sqrt{\frac{24\pi}{7}\frac{E_{\rm tot,min}}{V}} \, ,
\end{equation}
where $E_{\rm tot,min}/V=2.5\times10^{41} (1+k)^{4/7}P_{\rm1.4~GHz}^{4/7}V^{3/7}$, and $P_{\rm1.4~GHz}$ the radio power of the radio galaxy at 1.4 GHz, \citep[e.g.,][]{govoni+feretti04}.

Our results are reported in Table \ref{tab:nu_br}. The radio galaxies span a power range of $22<\log P_{\rm 1.4~GHz~[W~Hz^{-1}]}< 25$.
The most powerful objects are associated with the 
BCGs of the A\,3528 complex, while the weakest are associated with the radio galaxies belonging
to the A\,3558 complex. 
The equipartition values for the magnetic field and internal pressure are typical of radio galaxies 
in clusters  \citep[e.g.,][]{feretti+92,parma+99}. The internal 
pressure is higher for the most powerful radio galaxies.
The radiative ages for the central regions of the sources in our sample are on average of the order
of few times 10$^7$ yr, in agreement with literature works  \citep[e.g.,][]{murgia+11}. 

For the three BCGs  J\,1254-2913 (A\,3528~S), J\,1257-3021 (A\,3532) and 
J\,1324-3140 (A\,3556) our observations show the presence of extended emission with such 
steep spectrum which seems difficult to reconcile with the active region. In all cases the steep
spectrum radio emission extends well outside the optical counterpart.

\section{The host galaxies}\label{sec:ottico}
The optical (VST OmegaCAM $g$ band) images of the host galaxies are shown in Fig.~\ref{fig:morpho}. The visual inspection of the images is
sufficient to reveal the early-type nature of all galaxies in the sample. In most of them, the surface brightness declines smoothly from
the centre to the outskirts, with no sign of spiral structure or star-formation knots. There are however two galaxies showing evidence
of interaction, e.g. J\,1333-3140 (BCG in A\,3562) and J\,1254-2913 (BCG in A\,3528~S).

In the outskirts of J\,1333-3140, a ``ripple'' (or
``shell'', see \cite{B13}) is clearly visible in the N-W direction at
$\sim$25-30\,arcsec (i.e., $\sim$25-30\,kpc at galaxy redshift)
indicated by the white arrows in the last panel in Fig. ~\ref{fig:morpho} and in Fig. \ref{fig:optical_shell}.
A faint fan-shaped brightness features is visible on the opposite
side, plus a clear and a low surface brightness area extending
$\sim$80\,arcsec to the E.
All these features have the same colours and form a continuous
brightness distribution with the rest of the galaxy (see Fig. \ref{fig:optical_shell}), indicating that
they consist of the same stellar populations. Such kind of features
are typical of the last phases of a merger \citep[e.g.,][and references therein]{B13,bienayme+85}.

\begin{figure*}
\centering
\includegraphics[width= \textwidth]{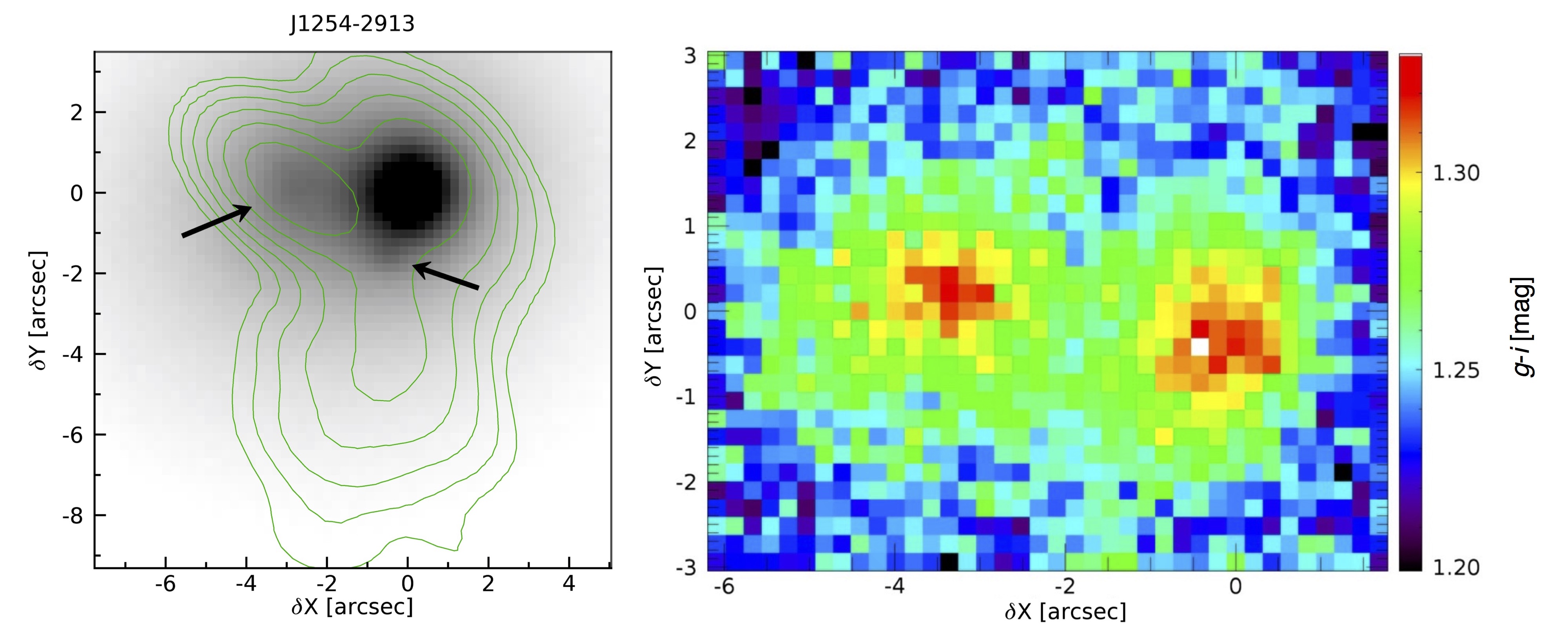}
\caption{{\it Left panel:} VST-OmegaCAM $i$-band image of the BCG J1254-2913 with the 8.4\,GHz radio contours overlaid (green). The left arrow points to a second nucleus, while the right arrow indicates a possible jet. {\it Right panel:} $g$-$i$ color map of the inner $8^{\prime\prime}\times6^{\prime\prime}$ of J1254-2913. Notice that the $g$-$i$ colors in the main nucleus of the galaxy (right in the figure) and in the secondary nucleus are similar and have similar gradients.}
\label{fig:J1254-2913_morpho}
\end{figure*}

An enlarged view of the central region of J\,1254-2913
is shown in the left panel in Fig.~\ref{fig:J1254-2913_morpho}. A round-shaped, faint secondary maximum of surface
brightness is visible about 3\,kpc East from the centre of the galaxy.
Another faint feature protrudes for about 2\,kpc South-South East from
the same centre.
To investigate the nature of the secondary maximum, we show in
the right panel of Fig.~\ref{fig:J1254-2913_morpho} the $g$-$i$ color distribution in the inner $8^{\prime\prime}\times6^{\prime\prime}$
of J\,1254-2913.
It is apparent that the $g$-$i$ color and the smooth color gradient of the
secondary maximum are similar to the nucleus of the galaxy, while the
range of values (1.2-1.34\,mag) is typical of early-type galaxy
populations \citep[e.g.,][]{gavazzi+10}. The secondary maximum is
therefore the nucleus of an early-type galaxy.
The color of the southern protrusion cannot be measured because of the
low SNR. Both features are accurately traced by the 8.4 GHz VLA
contours, suggesting that they could belong to the same galaxy. Due to
their low surface brightness, we were not able to measure their
redshift, hence the possibility of background/foreground objects,
though unlikely, cannot be excluded.
The presence of a double nucleus, together with the almost
unperturbed overall structure may indicate the final (pre-coalescence)
phase of a minor merger \citep[e.g.,][]{gimeno+04,lotz+10,mezcua+14},
in which the nucleus of the secondary, less massive
galaxy is sinking in the host potential well through dynamical
friction.

All the BCGs, with the exception of J\,1333-3140 (in A\,3562), have been observed and classified as cD galaxies by \citet{FBA10}. 
The homogeneous wavelength coverage available for all the galaxies in both complexes comprises the $gi$ optical filters from VST and the
WISE channels W1-W3 (at 3.4, 4.6 and 12\,$\mu$m). The corresponding magnitudes, corrected for the Galactic extinction, are listed in
Table~\ref{tab:mags}. The optical magnitudes were obtained with Sextractor \citep{BA96}. The values in Table~\ref{tab:mags} are Kron
magnitudes, which correspond to about 90\% of the total flux from the galaxy. For extended objects like our galaxies an accurate derivation
of the WISE magnitudes is crucial. The WISE magnitudes for extended objects ({\it gmag} in the WISE archive) are measured within the 2MASS
$K$-band isophototal aperture at 20\,mag\,arcsec$^{-2}$. There are though two 2MASS estimates of the total flux from extended sources,
given in the Extended Source Catalogue ($k_{\rm m_{k20fe}}$ and $k_{\rm m_{ext}}$), the second of which is based on extrapolation of the growth
curves. These extend out to different radii. 
We adopted the WISE magnitude measured in the aperture closer to that
used for the $i$-band magnitude. If it is the $k_{\rm m_{ext}}$ magnitude (curve of growth), then we offset all the WISE magnitudes by the difference in the two
2MASS $K$-band magnitudes ($k_{\rm m_{k20fe}} - k_{\rm m_{ext}}$).

\begin{figure}
\includegraphics[width=0.5\textwidth]{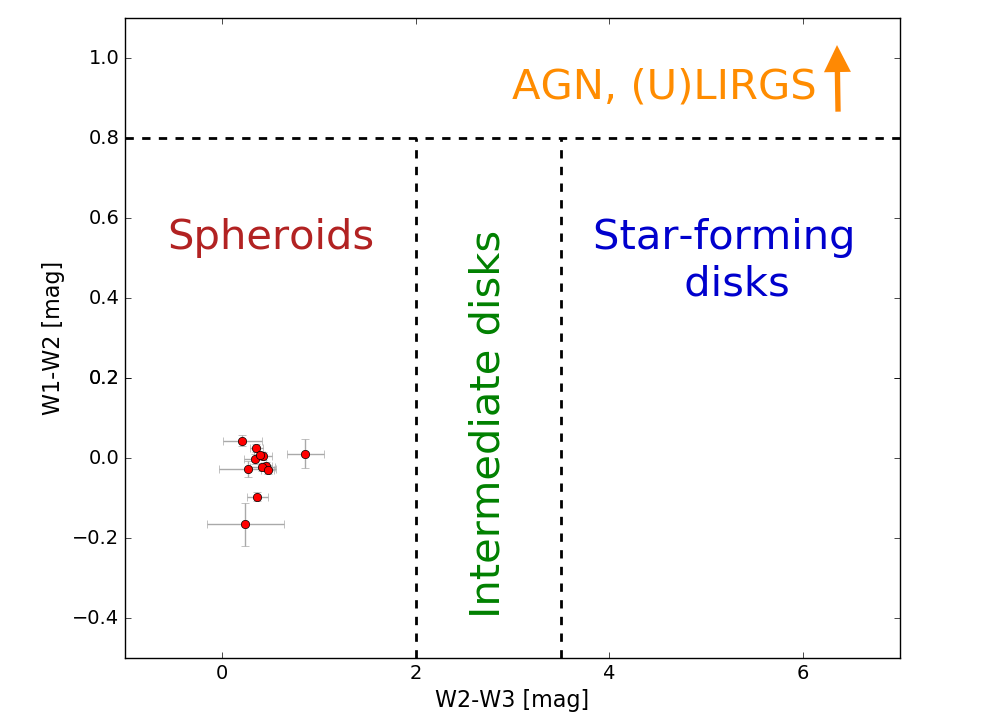}
\caption{Position of the 12 radio galaxies (dots with error bars) in
  the WISE color-color diagram. The diagram is divided in regions of
  colors characterizing different galaxy types following
  \citet{JCM17} (cf. their Fig. 11).}
\label{fig:WISE_colours}
\end{figure}

Fig.~\ref{fig:WISE_colours}, drawn from \citet{JCM17}, shows the WISE colors ranges characterizing different galaxy types according to
their star-formation or nuclear activity. The colors of all galaxies in our sample are well within the area relevant to the spheroidal
passive galaxies, as also indicated by their morphology.
We used the fluxes in the wavebands of Table~\ref{tab:mags} to derive the stellar masses for the galaxies (last column in the table) with
the spectral energy distribution (SED) fitting technique using the code \texttt{MAGPHYS} \citep{CCE08}. The code adopts the \citet{BC03} library of
models with a \citet{C03} stellar initial mass function and a metallicity value in the range $0.022-1~Z_{\odot}$. The age of the
galaxy is free to vary with an upper limit imposed by the age of the Universe at the considered redshift. In order to find the best-fit
model, \texttt{MAGPHYS} uses a Bayesian approach. As outputs, it produces the probability distribution function of all the parameters and their
best-fit values. For the stellar masses we adopted the median value as the more robust estimate. We run \texttt{MAGPHYS} using a library of templates
heavily biased to fit old stellar populations with very little star formation (kindly provided by E. da Cunha), since \texttt{MAGPHYS}, when run on
its public library, tends to favour dusty SF models rather than passive models \citep[e.g.][]{ABM14}.

\begin{table*}
\caption{Photometric properties and stellar masses of the host galaxies.}             
\label{tab:mags}      
\centering
{\small          
\begin{tabular}{l l l cccccc} 
\hline\hline\noalign{\smallskip}
Object & RA(J2000) & DEC(J2000) & $g$ & $i$ & W1 & W2 & W3 & $\log(M_\star/M_\odot$) \\ 
\hline\noalign{\smallskip}
{\bf A\,3528\,N} &&&&&&&& \\
J1254-2900 & 12\,54\,22.24 & -29\,00\,46.73 & 14.09$\pm$0.01 & 12.85$\pm$0.01 & 10.07$\pm$0.01 & 10.09$\pm$0.01 &  9.65$\pm$0.05 & 11.75$\pm$0.08 \\
J1254-2901a & 12\,54\,23.22 & -29\,01\,03.05 & 15.89$\pm$0.01 & 14.76$\pm$0.01 & 12.46$\pm$0.02 & 12.45$\pm$0.03 & 11.60$\pm$0.19 & 10.84$\pm$0.10 \\
J1254-2901b & 12\,54\,40.75 & -29\,01\,48.55 & 16.23$\pm$0.01 & 15.12$\pm$0.01 & 12.63$\pm$0.01 & 12.66$\pm$0.02 & 12.40$\pm$0.30 & 10.71$\pm$0.09 \\
J1254-2904 & 12\,54\,20.41 & -29\,04\,08.67 & 15.87$\pm$0.01 & 14.76$\pm$0.01 & 12.39$\pm$0.01 & 12.35$\pm$0.01 & 12.15$\pm$0.20 & 10.85$\pm$0.10 \\
{\bf A\,3528\,S} &&&&&&&& \\
J1254-2913 & 12\,54\,41.10 & -29\,13\,39.58 & 13.46$\pm$0.01 & 12.33$\pm$0.01 & 9.81$\pm$0.01 & 9.78$\pm$0.01 &  9.44$\pm$0.07 & 11.89$\pm$0.09 \\
J1254-2916 & 12\,54\,52.42 & -29\,16\,16.80 & 14.88$\pm$0.01 & 13.72$\pm$0.01 & 11.22$\pm$0.01 & 11.25$\pm$0.01 & 10.84$\pm$0.13 & 11.20$\pm$0.09 \\ 
{\bf A\,3532} &&&&&&&& \\
J1257-3021 & 12\,57\,21.94 & -30\,21\,49.00 & 14.34$\pm$0.01 & 13.27$\pm$0.01 & 10.87$\pm$0.01 & 10.87$\pm$0.01 & 10.54$\pm$0.11 & 11.43$\pm$0.10 \\   
{\bf A\,3556} &&&&&&&& \\
J1324-3138 & 13\,23\,57.61 & -31\,38\,44.82 & 14.69$\pm$0.01 & 13.46$\pm$0.01 & 10.91$\pm$0.01 & 10.90$\pm$0.01 & 10.49$\pm$0.09 & 11.38$\pm$0.08 \\
J1324-3140 & 13\,24\,06.73 & -31\,40\,11.56 & 13.89$\pm$0.01 & 12.74$\pm$0.01 & 10.01$\pm$0.01 & 10.00$\pm$0.01 & 9.62$\pm$0.06 & 11.66$\pm$0.07 \\  
{\bf A\,3558} &&&&&&&& \\
J1327-3129b & 13\,27\,56.86 & -31\,29\,45.32 & 13.29$\pm$0.01 & 12.17$\pm$0.01 & 9.39$\pm$0.01 & 9.49$\pm$0.01 & 9.13$\pm$0.11 & 11.85$\pm$0.08 \\  
{\bf A\,3562} &&&&&&&& \\
J1333-3140 & 13\,33\,34.73 & -31\,40\,20.34 & 14.13$\pm$0.01 & 12.97$\pm$0.01 & 10.77$\pm$0.01 & 10.81$\pm$0.01 & 10.34$\pm$0.07 & 11.46$\pm$0.09 \\
J1333-3141 & 13\,33\,31.63 & -31\,40\,58.33 & 16.30$\pm$0.01 & 15.09$\pm$0.01 & 12.63$\pm$0.02 & 12.80$\pm$0.05 & 12.57$\pm$0.39 & 10.70$\pm$0.08 \\
\hline
\end{tabular}
} 
\tablefoot{Columns 2 and 3: galaxy coordinates corresponding to the photometric centre in $g$ band. Columns 4 and 5: optical ($gi$) magnitudes from ESO-VST 
given in AB photometric system \citep[see][]{merluzzi+15,mercurio+15}. Columns 6 to 8: mid-infrared magnitudes from WISE channels W1, W2 and W3 derived as 
explained in the text. All magnitudes are corrected for the Galactic extinction following \citet{SF11}. Column 9: logarithm of stellar mass in units of solar mass.}
\end{table*}

We find that the mean stellar masses of the BCGs are $\langle\log(M_\star/M_{\odot})\rangle=11.66\pm0.09$ in the A\,3558 complex and
$\langle\log(M_\star/M_{\odot})\rangle=11.69\pm0.10$ in the A\,3528 complex. The mean stellar masses of the tailed galaxies are
$\langle\log(M_\star/M_{\odot})\rangle=10.09\pm0.09$ and $\langle\log(M_\star/M_{\odot})\rangle=11.04\pm0.24$ in the A\,3558 and A\,3528
complex, respectively.

\section{Discussion}\label{sec:disc}
We present a detailed radio analysis of the BCGs and tailed radio galaxies in the  A\,3528 and 
A\,3558 cluster complexes, to investigate whether the large scale environment is influencing
the emission and the cycles of activity of the radio galaxies. 

Our study was based on proprietary and literature high-sensitivity radio observations, obtained with the GMRT at 235, 325 and 610 MHz;  MOST at 843 MHz; VLA at 1.4 and 8.4 GHz; ATCA at 1.38, 2.38 and 5 GHz. The multifrequency radio images show that the radio properties of the brightest cluster galaxies in the A\,3528 and A\,3558 complex are very different, as detailed below.

\subsection{The BCGs}\label{sec:bcgs}
The two BCGs located at the centre of the main clusters in the two complexes, i.e., J\,1254-2900 
in A\,3528~N and J\,1327-3129b in A\,3558, are both embedded in the faint optical halo emission, 
but their radio properties are completely different. 
The former has a clear mini-double morphology (linear size of about 30 kpc) and its spectrum 
is well modeled by a power law, with a spectral index typical of an active source 
($\alpha\approx0.8$), implying a young population of relativistic electrons. Moreover, it is powerful 
in the radio band ($\log P_{\rm 1.4~GHz~[W~Hz^{-1}]}\approx24$). On the contrary, J\,1327-3129b (A\,3558) is compact, 
small (linear size of $\approx20$ kpc) and very weak ($\log P_{\rm 1.4~GHz~[W~Hz^{-1}]}\approx22.5$). 
Its spectrum is concave and is suggestive of
a compact steep-spectrum (CSS) radio galaxy, i.e. a radio galaxy in the very early stages of its evolution. The very steep spectrum in the optically thin part of the spectrum further suggests the presence of a population of old electrons.
Short {\it Chandra} observations (ObsID 1646) have also revealed the presence of a cool corona  in the 1-2 keV band surrounding the optical and radio emission \citep{sun09}.
Overall, the radio and X-ray properties of J\,1327-3129b suggest that the BCG in A\,3558 is an old CSS, whose radio emission is completely confined by the X-ray corona. The study of the interplay between the radio emission of the galaxy and the ICM will be presented in a future work.

We postulate that the differences of the radio properties of these two BCGs are related to the 
different environment in which they are located. Previous studies of the A\,3528 complex 
\citep{gastaldello+03} suggest that the individual clusters have not crossed each other, and 
their nuclear regions have not yet undergone major disruption. This is  confirmed by the presence of cool-cores \citep[see Fig. 7 in][]{gastaldello+03}, an indication that is a relaxed system.
On the other hand,  multi-band observations provide support to the idea that the A\,3558 complex  has undergone a major merger, particularly active in the region between A\,3562 and A\,3558
\citep[e.g.,][]{bardelli+98a,bardelli+98b,merluzzi+16,ettori+00,rossetti+07,ghizzardi+10,venturi+00,venturi+03,giacintucci+04,giacintucci+05,venturi+17}.  Considering that the BCG in A\,3562 is radio quiet, and the well-known merging state from literature studies, our result is another piece of evidence suggesting
that another observational implication of major mergers is the quenching of the nuclear radio emission in the 
BCGs \citep{kale+15}. Despite that, we are aware of the fact that the interplay between the cluster dynamics and the radio emission properties of cluster
galaxies is not fully understood yet, and that a number of BCGs are also found radio quiet in relaxed environments \citep{sun09}.

The remaining BCGs in our sample, i.e. J\,1254-2913 (A\,3528~S), J\,1257-3021 (A\,3532) and 
J\,1324-3140 (A\,3556), are  similar in their global properties. 
They are all powerful in the radio band ($\log P_{\rm 1.4~GHz~[W~Hz^{-1}]}\approx23-25$). 
They are characterised by 
a compact active region and by diffuse emission, which extends well beyond the optical counterpart,
and has a radio spectrum much steeper than the nuclear component.
Their overall morphology, the clear separation between the active nucleus and the diffuse
emission, and the much steeper spectrum of the latter suggest that these 
objects could be examples of re-started radio sources. 

We point out that a number of restarted radio galaxies have been found at the centres of relaxed 
groups and clusters \citep[e.g.,][]{giacintucci+07,murgia+11,giacintucci+14a,shulevski+15}, as it is indeed the case of  A\,3528~S, A\,3532 and A\,3556. Indeed the latter does not seem to be involved in the major merging occurring in the supercluster core (see above).

We do not observe such differences in the optical properties of the host galaxies of the BCGs in the two cluster complexes (Fig.~\ref{fig:WISE_colours}). They are all passive, belong to the red-sequence, do not show relevant signs of (un)obscured star formation and have on average the same high stellar masses ($\langle \log(M_\star/M_\odot) \rangle =11.33$). 
We speculate that cluster mergers do not affect the host galaxies, at least not on the life time-scale
typical of the radio emission, i.e. of the order of  $10^7 - 10^8$ years \citep[i.e., $1.4-7.8\times10^7$ years; see][]{murgia+03,murgia+11}.

\subsection{Tails and cluster weather in A\,3528}\label{sec:merg_dyn}
In contrast to the A\,3558 complex, where there is substantial observational support in favor of
the merger and accretion activity, little is known about the A\,3528 complex, in particular 
concerning the two sub-cluster A\,3528~N and A\,3528~S. Literature studies in the optical band \citep{bardelli+01} show that the galaxy velocity distribution is characterized by the presence of several peaks \citep[see Fig. 4a in][]{bardelli+01}, which is suggestive of substructure, and possibly indicating some sort of interaction between the two sub-clusters. Despite that, \citeauthor{bardelli+01} also found that this velocity distribution is well approximated by a Gaussian, which implies that any interaction between these two sub-clusters, if indeed present, has not affected the observed radio and optical properties of the galaxy population yet. 
Along this interpretation, the thermal properties of the ICM \citep{gastaldello+03} show a significant excess in the  {\it XMM-Newton} surface brightness 
northwest of A3528~N and northeast of A3528~S. This information has been interpreted as 
a scenario in which the two systems are orbiting around each other, as a consequence of an 
off-axis post-(minor)merger event.

The presence of several tailed radio galaxies in the A\,3528 complex is a valuable addition
to the available information on the so far unclear dynamical state of the A\,3528 complex. 
Three tails, i.e. J\,1254-2901a and J\,1254-2901b in A\,3528~N and J\,1254-2916 in A\,3528~S, are all suggesting an eastward
motion, at least on the plane of the sky. The S-shaped morphology of the small double radio galaxy associated with
the BCG in A\,3528 is a further suggestion of some form of dynamical activity
in this cluster \citep{gopal-krishna+12}.

The tailed radio galaxy J\,1254-2904 in A\,3528~N) is intriguing: it is located between A\,3528~N and A\,3528~S, suggesting
a galaxy motion towards North, and presents a peculiar bottle-neck morphology possibly due to integration with the ICM. On the basis of {\it XMM-Newton} observations, \cite{gastaldello+03} proposed indeed a scenario in which A\,3528~N and A\,3528~S are moving along the N-W/S-E and N-E/S-W direction, respectively, with a merger occurred about 1-2 Gyrs ago that was sufficient to disturb only the outer galaxy distribution but not the cluster cores. 
Recent numerical simulations \citep{jones+17} and new observational results \citep[e.g.][]{cuciti+17,degasperin+17,digennaro+18} have shown that 
if the jet plasma is impacted perpendicularly by wind (i.e. ``cross wind'') resulting from a crossing shock, it
could be displaced and revived, resulting into the broken-shape morphology we see for this source. Although spectral index analysis across the tail is necessary to confirm such scenario, no clear emission from the tail of the radio galaxy is provided at high frequencies ($>610$ MHz). Another less exciting possibility
is that the southern region of emission is actually associated with
the galaxy at z $\sim$ 0.07 shown in Fig. \ref{fig:opt-radio_tailA3528}. Future analysis will be provided with our new upcoming uGMRT observations at 1.4 GHz.

The optical analysis of the host galaxies of the tailed radio sources does not show any difference, either between the two cluster complexes and with the BCGs discussed above.

\section{Conclusions}\label{sec:conc}
Our study of the radio properties of the BGCs and tailed radio galaxies in the central region of
the Shapley Concentration provides information on different aspects of galaxy clusters
 and their formation. While the radio properties of the BCGs provide information on the activity state in the central dense environment of the galaxy clusters, tailed radio galaxies are tracers of the motion of the host galaxies within the clusters, which take into account also the bulk flow motion in the cluster weather. The main conclusions of this paper can be summarised as follow:

\begin{itemize}
\item[$\bullet$] Our 235, 325 and 610 MHz radio images have shown different radio galaxy morphology contents for the two cluster complexes: while the A\,3558 one is dominated by compact radio sources (Fig. \ref{fig:mosaic_A3558}), the A\,3528 one shows the presence of several radio tails and diffuse radio emission surrounding the BCGs (i.e. J\,1254-2913 in A\,3528~S and J\,1257-3021 in A\,3532, Fig \ref{fig:mosaic_A3528}).
\medskip
\item[$\bullet$] We found powerful radio active BGCs in the A\,3528 complex. Two of them (J\,1254-2913 and J\,1257-3021, located in A\,3528~S and A3532 respectively) are suggestive of restarted activity, with an active nucleus (as clear derived from its spectral properties) and radio emission with very steep spectrum extending out in the ICM well beyond the boundaries of the optical counterpart and possibly tracing a previous cycle of activity. Interestingly, the broad band properties of these clusters suggest that they are either relaxed, or have not undergone a disruptive merger event.
\medskip
\item[$\bullet$] 
The BCGs in A\,3558 (i.e. J\,1327-3129b) and in A\,3562 (i.e. J\,1333-3141) show remarkable differences from the ones in the A\,3528 complex. The former is a faint compact source with concave spectrum, typical of compact steep spectrum sources. The slope in the optically thin part of the spectrum is quite steep, suggesting that this source has already aged and will not evolve in an extended source, i.e., the radio plasma may not expand beyond its present size and the radio source may evolve confined by the external medium. The latter is radio quiet at the sensitivity limits of our observations. Despite such differences, considering the several pieces of observational evidence in favour of the fact that A\,3558 and A\,3562 have undergone a recent major merger event, it is tempting to conclude that this has severely affected the radio emission of the BCGs.
\medskip
\item[$\bullet$] The BCG in A\,3556 (i.e. J\,1324-3138), though less powerful, exhibit properties comparable  with those of the A\,3528 complex, with hints of restarted activity, i.e., it has a compact core whose flat spectrum is suggestive of variability, and two faint lobes which are well imaged only at 325 MHz and below. A\,3556 is at the extreme western periphery of the A\,3558 complex and the available multiband information suggests that is has not taken part in the major merger event of the A\,3558-A\,3562 region.

\smallskip
\item[$\bullet$] We performed a study on the host galaxies in both complexes. Not surprisingly, they are associated with high-stellar mass passive galaxies. We do not observe any difference between the A\,3528 and A\,3558 complexes. We speculate that the cluster dynamical state does not affect the optical counterparts of the radio galaxies, at least on the life-time of the radio emission, i.e., $10^7-10^8$ yr.
\end{itemize}

A detailed study of the interplay between the diffuse radio emission from the BCGs and the X-ray emission from the ICM, aimed to confirm the hypothesis of restarted radio activity and the presence of X-ray cavities/coronae, is in progress and will be presented in a future work. In particular, we will make use of our new uGMRT observations.

\medskip

\begin{acknowledgements}
We thank the anonymous referee for carefully reading the paper, and for the useful comments and suggestions which have improved the clarity of the manuscript and
the presentation of our results.
This paper is based on data obtained with the Giant Metrewave Radio Telescope (GMRT) and the Very Large Array (VLA).
GMRT is run by the National Centre for Radio Astrophysics of the Tata Institute of Fundamental Research. The National Radio Astronomy Observatory is a facility of the National Science Foundation operated under cooperative agreement by Associated Universities, Inc.
The VLA is operated by the National Radio Astronomy Observatory, which is a facility of the National Science Foundation operated under cooperative agreement by Associated Universities, Inc. This research has made use of the NASA/IPAC Extragalactic Database (NED) which is operated by the Jet Propulsion Laboratory, California Institute of Technology, under
contract with the National Aeronautics and Space Administration. 
Basic research in radio astronomy at the Naval Research Laboratory is supported by 6.1 Base funding.
CPH acknowledges financial support from PRIN INAF 2014. PM, GB and AM acknowledge funding from the INAF PRIN-SKA 2017 program 1.05.01.88.04. The authors thank E. Iodice who collected the g-band imaging of the A\,3528 complex at ESO-VST.
\end{acknowledgements}

\appendix

\section{Optical images of the host galaxies} 

\begin{figure*}[h!]
\centering
\includegraphics[width=0.9\textwidth]{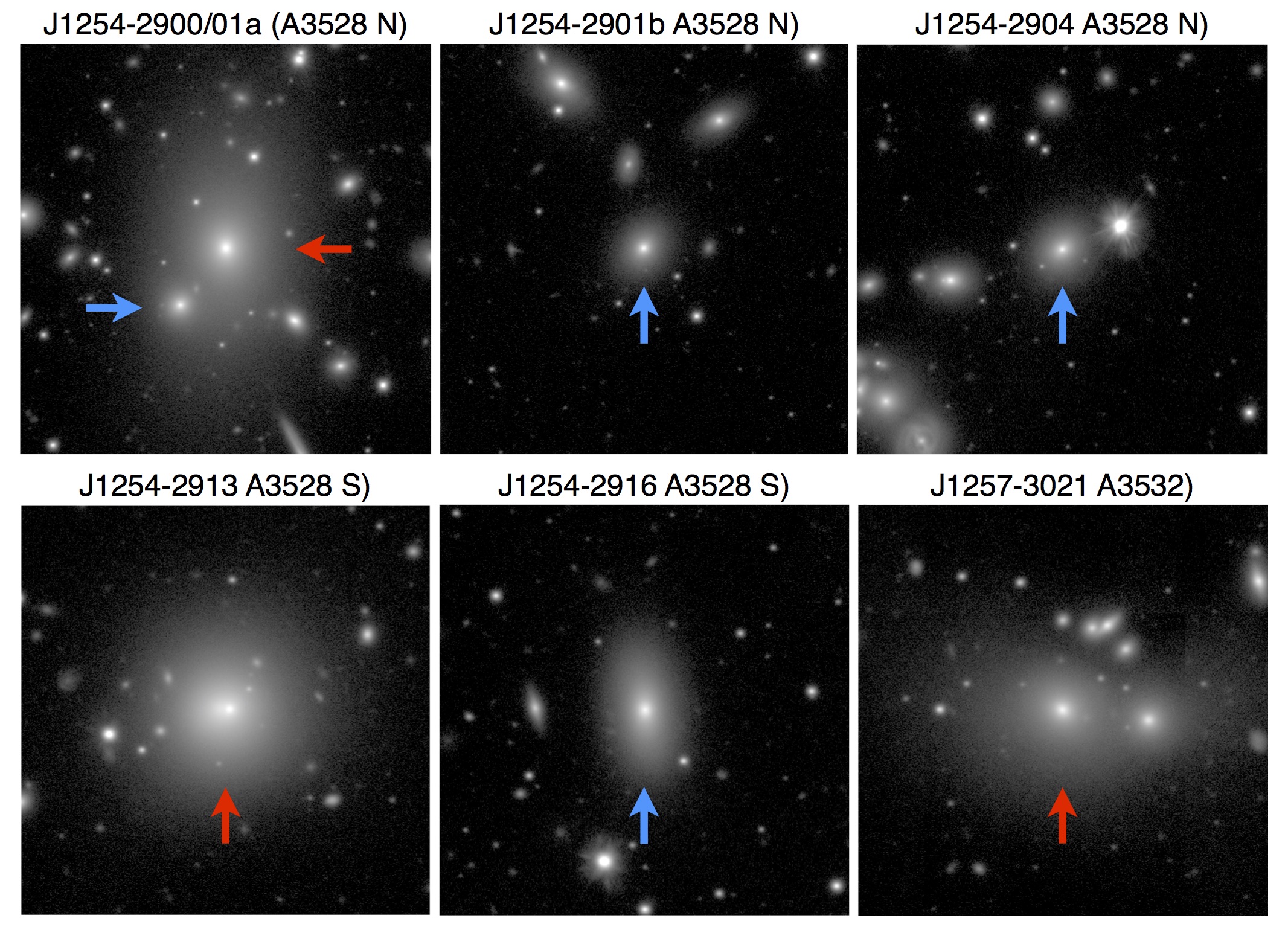}
\includegraphics[width=0.9\textwidth]{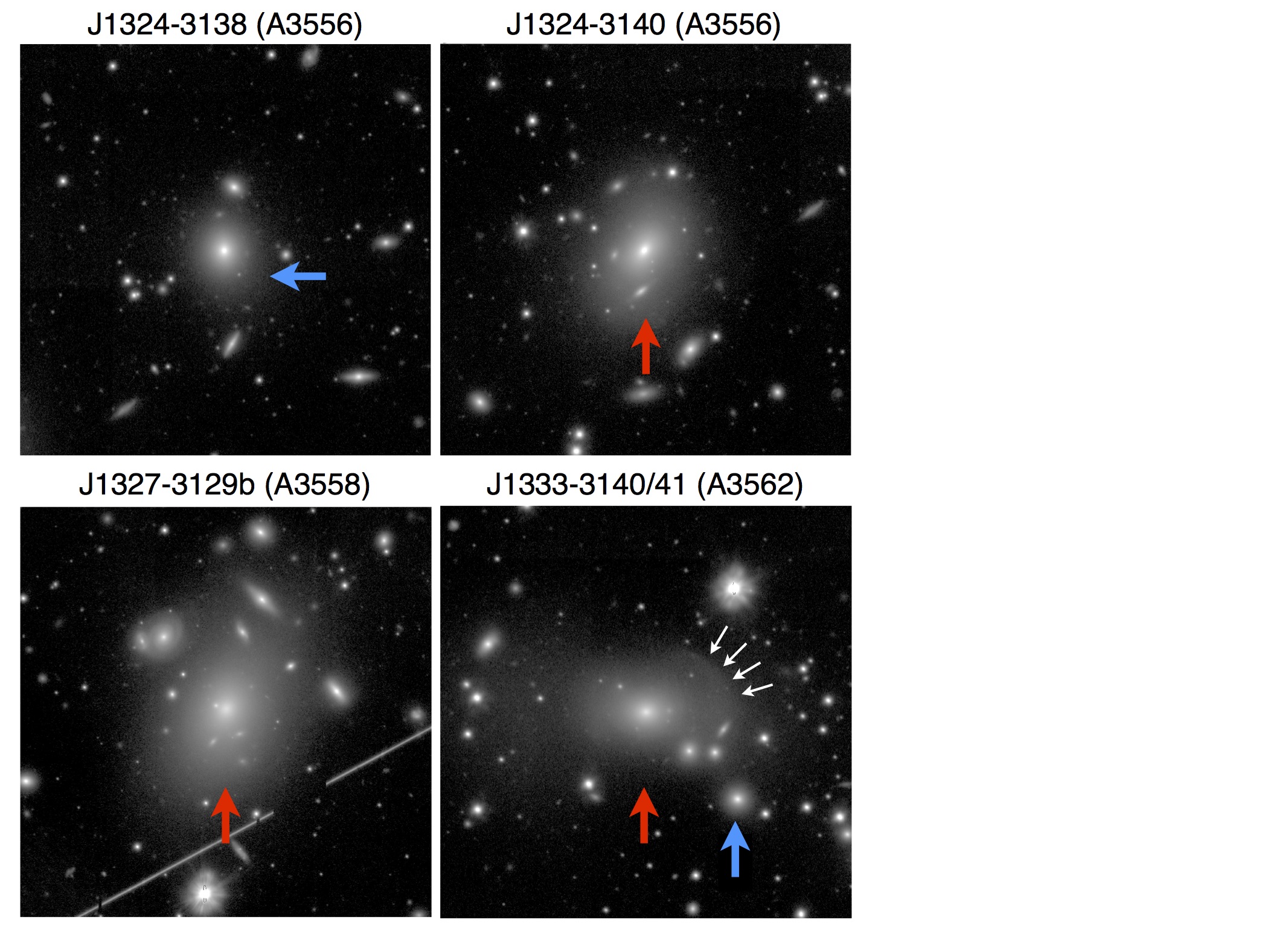}
\caption{VST-OmegaCAM $g$-band images of the host galaxies (figure center). Each panel
  has 2\,arcmin sides. North is up and West to the right. Galaxies
  ID(s) are indicated. The six top panels show the galaxies in the
  A\,3528 complex, while the four bottom panels show galaxies in the
  A\,3558 complex. Red and blue arrows indicate the BCGs and the tailed radio galaxy, respectively, with the same color code as for Fig. \ref{fig:mosaic_A3528} and \ref{fig:mosaic_A3558}. 
  Withe arrows in the last right bottom panel represent the position of the optical shell.}
\label{fig:morpho}
\end{figure*}

\begin{figure}[ht]
\centering
\includegraphics[width=0.5\textwidth]{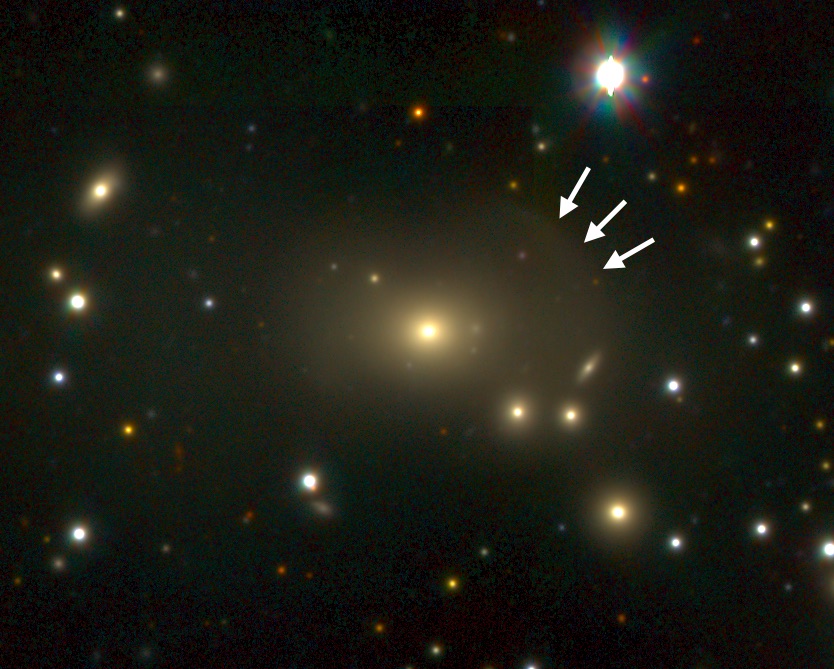}
\caption{VST-ACESS {\it gri} composite images of the BCG in A\,3562 which shows the similar brightness distribution from the shell towards the optical cD position. As for the last panel in Fig.~\ref{fig:morpho}, the white arrows highlights the position of the optical shell.}\label{fig:optical_shell}
\end{figure}


\bibliography{biblio.bib}

\end{document}